\pdfmapfile{=<name>.map}
\documentclass[10pt,sigconf]{acmart}

\AtBeginDocument{%
  \providecommand\BibTeX{{%
    \normalfont B\kern-0.5em{\scshape i\kern-0.25em b}\kern-0.8em\TeX}}}

\usepackage{balance}  %
\usepackage{graphics} %
\usepackage{url}      
\usepackage{graphicx}
\usepackage{tabularx}
\usepackage{float}
\usepackage{xcolor}
\usepackage{colortbl}
\usepackage{url}
\usepackage[noend]{algpseudocode}
\usepackage{algorithm}
\usepackage{verbatim}
\usepackage{mathtools}
\usepackage{amsthm}
\usepackage{xspace}
\usepackage{mathrsfs}
\usepackage{listings}
\usepackage{color}
\usepackage{booktabs}
\usepackage{esvect}
\usepackage{subfigure}
\usepackage{makecell}
\usepackage{multirow}
\usepackage{tikz}
\usepackage{relsize}
\usepackage{capt-of}
\usepackage{enumitem}
\usepackage{dcolumn}\newcolumntype{d}[1]{D{.}{.}{#1}}
\usepackage[compress]{cleveref}
\usepackage[normalem]{ulem}
\usepackage[defaultlines=3,all]{nowidow}

\newcommand{\squeeze}[1]{\textls[-10]{#1}}
\newcommand{\squeezemore}[1]{\textls[-20]{#1}}
\newcommand{\squeezeevenmore}[1]{\textls[-30]{#1}}

\newcommand*\circled[1]{\tikz[baseline=(char.base)]{
    \node[shape=circle,draw,fill=orange,inner sep=0pt] (char) {#1};}}

\makeatletter
\newcommand{\eqnum}{\leavevmode\hfill\refstepcounter{equation}\textup{\tagform@{\theequation}}}
\makeatother

\makeatletter
\newcommand*{\bdiv}{%
  \nonscript\mskip-\medmuskip\mkern5mu%
  \mathbin{\operator@font div}\penalty900\mkern5mu%
  \nonscript\mskip-\medmuskip
}
\makeatother

\newcommand{\avc}{\textsc{h264}\xspace}
\newcommand{\hevc}{\textsc{hevc}\xspace}

\newcommand{\yuvfourtwozero}{\textsc{yuv420}\xspace}
\newcommand{\yuvfourtwotwo}{\textsc{yuv422}\xspace}
\newcommand{\rgb}{\textsc{rgb}\xspace}

\newcommand\CC{C\nolinebreak[4]\hspace{-.05em}\raisebox{.7ex}{\relsize{-1}{\textbf{++}}}\xspace}

\crefname{section}{Section}{Sections}
\Crefname{section}{Section}{Sections}
\crefname{figure}{Figure}{Figures}
\Crefname{figure}{Figure}{Figures}
\crefname{subfigure}{Figure}{Figures}
\Crefname{subfigure}{Figure}{Figures}
\crefrangelabelformat{subfigure}{#3#1#4--#5(\crefstripprefix{#1}{#2}#6}

\newcommand{\todo}[1]{\textcolor{red}{\{\{#1\}\}}}

\newcommand{\alvin}[1]{\textcolor{magenta}{[[Alvin: #1]]}}

\newcommand{\revisioncolor}{black}
\newcommand{\shepherdcolor}{black}
\newcommand{\revision}[1]{{\color{\revisioncolor} #1}}
\newcommand{\shepherd}[1]{{\color{\shepherdcolor} #1}}

\makeatother\theoremstyle{definition}

\newtheoremstyle{definitions}%
  {\topsep}{\topsep}{\normalfont}{}%
  {\bfseries}{.}{5pt}{}
\theoremstyle{definitions}

\newtheoremstyle{solution}%
  {\topsep}{\topsep}{\normalfont}{}%
  {\bfseries}{.}{5pt}{}
\theoremstyle{solution}

\newtheoremstyle{claim}%
  {\topsep}{\topsep}{\normalfont}{}%
  {\bfseries}{.}{5pt}{}
\theoremstyle{claim}

\setcellgapes{5pt}
\acmConference[]{}{}{}%
\copyrightyear{2021}
\acmYear{2021}

\setcopyright{none}
\renewcommand\footnotetextcopyrightpermission[1]{}
\newcommand{\name}{VSS\xspace}

\begin{document}
\fancyhead{}

\title{VSS: A Storage System for Video Analytics} %
\subtitle{Technical Report}

\begin{sloppypar}

\begin{abstract}

We present a new video storage system (\name) designed to decouple high-level video 
\squeezemore{
operations from the low-level details required to store and efficiently retrieve video data.  
\revision{\name is designed to be the storage subsystem of a  video data management system (VDBMS)} and is responsible for:
(1) ~transparently and automatically arranging the data on disk in an efficient, granular format; (2)~caching frequently-retrieved regions in the most useful formats; and (3)~eliminating redundancies found in videos captured from multiple cameras with overlapping fields of view.  Our results suggest that \name can improve VDBMS read performance by up to 54\%, reduce storage costs by up to 45\%, and enable developers to focus on application logic rather than video storage and retrieval.
}

\end{abstract}

\iftrue
\author{Brandon Haynes*, Maureen Daum$^\dagger$, Dong He$^\dagger$, Amrita Mazumdar$^\dagger$}
\author{Magdalena Balazinska$^\dagger$, Alvin Cheung$^\ddagger$, Luis Ceze$^\dagger$}
\affiliation{%
  \institution{* Gray Systems Lab, Microsoft (brandon.haynes@microsoft.com)}
  \institution{$^\dagger$ University of Washington (\{mdaum, donghe, amrita, magda, luisceze\}@cs.washington.edu)}
  \institution{$^\ddagger$ University of California, Berkeley (akcheung@cs.berkeley.edu)}
}

\renewcommand{\shortauthors}{B. Haynes et al.}

\maketitle

\newcommand{\compactAlgorithm}{
\begin{algorithm}[t]
\caption{Contiguous materialized view compaction}
\label{alg:compact}
\begin{algorithmic}
  \Ensure All pairs of contiguous physical videos compacted
  \State\textbf{let} $\textsc{hardlink}(f, d)$ create hard link for file $f$ in dir $d$
  \State\textbf{let} $\textsc{insert}(m, f)$ insert file $f$ into the temporal index of $m$
  \State\textbf{let} $\textsc{dir}(m)$ be the directory associated with $m$
  \State\textbf{let} $\textsc{views}(v)$ be the physical videos associated with $v$
  \State\textbf{let} $\textsc{config}_{s,p}(m)$ give the spatial/physical config of $m$
  \State\textbf{let} $s_{m_i}, e_{m_i}$ be the start and end times of $m_i$
  \\
  \ForAll{$v_1 \in VSS, v_2 \in VSS \setminus v_1$}
    \ForAll{$m_1 \in \textsc{views}(v_1), m_2 \in \textsc{views}(v_2)$}
      \If{$e_{m_1}{=}s_{m_2} \land \textsc{config}_{s,p}(m_1) = \textsc{config}_{s, p}(m_2)$}
        \State \textbf{let} $F_2 = \textsc{dir}(m_2)$ sorted by time
        \ForAll{$f_2 \in F_2$}
          \State $\textsc{hardlink}(f_2, \textsc{dir}(m_1))$
          \State $\textsc{insert}(m_1, f_2)$    
        \EndFor

        \State $\textsc{delete}(m_2)$
      \EndIf    
    \EndFor
  \EndFor
\end{algorithmic}
\end{algorithm}
}

\newcommand{\budgetAlgorithm}{
\begin{algorithm}[t]
\caption{Storage budget algorithm}
\label{alg:budget}
\begin{algorithmic}
  \State \textbf{let} $\textsc{budget}(v)$ be the storage budget for logical video $v$
  \State \textbf{let} $\textsc{size}(m)$ be the actual size of $m$ on disk
  \State \textbf{let} $\textsc{estimate}(m)$ be the estimated size of $m$
  \State \textbf{let} $\textsc{physical}(m)$ be the physical videos for $v$
  \\
  \Function{admit}{$v\colon \text{logical video}, m\colon \text{physical video}$}
    \State \textbf{let} $M = \textsc{physical}(v)$
    \If{$\textsc{estimate}(m) + \sum_M \textsc{size}(m_i) \le \textsc{budget}(v)$}
        \State \textbf{return} \textsc{true}
    \Else
        \State \textbf{let} $x = \textsc{victim}(v, M \cup \{m\})$
        \If{$x \ne m$}{~$\textsc{prune}(v)$}
        \EndIf    
        \State $\textbf{return}~x \ne m \land \textsc{admit}(v, m)$
    \EndIf
  \EndFunction  

  \Function{prune}{$v\colon \text{logical video}$}
    \State \textbf{let} $M = \textsc{physical}(v) \setminus m_0$
    \While{$M \ne \emptyset \land \sum_M \textsc{size}(m_i) > \textsc{budget}(v)$}
    	\State $x \leftarrow \textsc{victim}(v, M)$
        \State $M \leftarrow M\setminus x$
    	\State $\textsc{delete}(x)$
    \EndWhile
  \EndFunction  
\end{algorithmic}
\end{algorithm}
}

\newcommand{\jointCompressionAlgorithm}{
\begin{algorithm}[t]
\caption{\revision{Joint compression algorithm}}
\label{alg:joint-compression}
\scriptsize
\color{\revisioncolor}
\begin{algorithmic}
  \State\textbf{let} $\textsc{homography}(f, g)$ estimate the $3{\times}3$ homography matrix of $f$ and $g$
  \State\textbf{let} $u, \tau, \epsilon$ respectively be the \name quality model (\autoref{sec:quality-model}), quality \\
  \hspace{4.25em} threshold (\autoref{sec:cost-model}), and duplicate threshold (\autoref{subsec:joint-compression-frame}).

  \vspace{0.5em}
  \Function{joint-compress}{$F, G, m$}
    \State \textbf{Input:} Video frames $F = \{f_1, ..., f_n\}$
    \State \textbf{Input:} Video frames $G = \{g_1, ..., g_n\}$
    \State \textbf{Input:} Merge function $m$
    \State \textbf{Output:} Vector of compressed subframes

    \vspace{0.5em}
    \State $H, C, i, j \leftarrow \textsc{homography}(f_1, g_1), \emptyset, 0, 0$

    \vspace{0.5em}
    \If{$H = \emptyset$} \textbf{return} $\emptyset$ \Comment{No homography found\phantom{00}}
    \ElsIf{$H_{1, 2} < 0$} \textbf{return} $\textsc{joint-compress}(G, F)$ \Comment{Reverse transform\phantom{00}}
    \EndIf

    \vspace{0.5em}
    \While{$i \le n$}
      \If{$\left\vert\vert H - \mathcal{I} \vert\right\vert_2 \le \epsilon$} 
        \State $S_i, H \leftarrow (\emptyset, f_1, \emptyset), \mathcal{I}$ \Comment{Duplicate Frames\phantom{00000ll}}
      \Else
        \State $S_i\leftarrow \textsc{partition}(f_i, g_i, H, m)$
      \EndIf

      \If{$S_i = \emptyset \lor u\Big(f_i, \big[S_i^{\text{left}}~S_i^{\text{overlap}}\big]\Big) < \tau~\lor$ \Comment{Verify quality\phantom{000000000l}}\\
          \hspace{4em}$u\Big(g_i, \big[\textsc{transform}(S_i^{\text{overlap}}, H^{-1})~S_i^{\text{right}}\big]\Big) < \tau$}
        \If{j = 0} \Comment{Recompute homography}
          \State $H, j \leftarrow \textsc{homography}(f_i, g_i), j + 1$ 
        \Else
          \State \textbf{return} $\emptyset$ \Comment{Abort joint compression}
        \EndIf
      \Else
        \State $C \leftarrow C \oplus \textsc{compress}(S_i)$
        \State $i, j \leftarrow i + 1, 0$
      \EndIf
    \EndWhile

    \State \textbf{return} $C$
  \EndFunction
  \\
  \Function{partition}{$f, g, H, m$}
      \State \textbf{Input:} Video frame $f$ with size $n, m$
      \State \textbf{Input:} Video frame $g$ with size $n, m$
      \State \textbf{Output:} Left, overlap, and right subframes

      \vspace{0.5em}
      \State $x_f \leftarrow {\big[ H^{-1} \cdot \begin{bmatrix}0 & 0 & 1\end{bmatrix}^\intercal \big]}_{2}$
      \State $x_g \leftarrow n - {\big[ H \cdot \begin{bmatrix}n & 0 & 1\end{bmatrix}^\intercal \big]}_{2}$

      \vspace{0.25em}
      \If{$\neg (0 < x_f \le n) \lor \neg (0 < x_g \le n)$} $\textbf{return}~\emptyset$\EndIf

      \vspace{0.25em}
      \State $l, r \leftarrow f[1, x_f], g[x_g, n]$
      \Comment{Left and right subframes}
      \State $o \leftarrow m(f[x_f, n], \textsc{transform}(g[1, x_g], H))$
      \Comment{Overlap of $f$ and $g$\phantom{00000}}
      \State \textbf{return} $(l, o, r)$
  \EndFunction
  \\
  \Function{transform}{$f, H$} \Comment{Compute perspective transform of $f$}
    \State \textbf{return} $\begin{bmatrix}f'_{1,1} & \hdots \\ \vdots & \ddots\end{bmatrix}$, where $t_{i,j} = H \cdot \begin{bmatrix}i\\ j\\ 1\end{bmatrix},~f'_{i, j} = f_{\frac{[t_{i, j}]_{1}}{[t_{i, j}]_{3}}, \frac{[t_{i, j}]_{2}}{[t_{i, j}]_{3}}}$
  \EndFunction
\end{algorithmic}
\end{algorithm}
}

\newcommand{\queryAnswerAlgorithm}{
\begin{algorithm}[t]
\caption{Query answering}
\label{alg:query-answer}
\begin{algorithmic}
  \State \todo{This hasn't been updated to conform with the text in Section 3 yet}
  \\
  \State \textbf{Input:} VSS query $scan(D, S_q, T_q=(s_q, e_q), P_q)$
  \State\textbf{let} $M=\{m_1, ..., m_n\}$ be the materialized views of $D$
  \State\textbf{let} $C(m, T)$ convert view $m$ to configuration $S_q, T, P_q$
  \State\textbf{let} $v_i \oplus v_j$ concatenate two materialized views

  \State \textbf{let} $V = \{s_q, e_q\}$
  \State \textbf{let} $E = A = \emptyset$
  \\
  \ForAll{$m_i \in M$}
    \State $V \leftarrow V \cup \{s_i, e_i\}$
  \EndFor

  \ForAll{$m_i \in M, m_j \in M\setminus m_i$}
    \If{$s_i < s_j \land e_i \ge s_j$}
      \State $E \leftarrow E \cup (s_i, e_j)$ \alvin{is $e_j$ a typo?}
    \EndIf  
    \If{$s_i < e_j \land e_i \ge e_j$}
      \State $E \leftarrow E \cup (s_i, e_j)$
    \EndIf  
  \EndFor

  \State \textbf{let} $G = (V, E)$
  \State \textbf{let} $P = \textsc{shortest-path}(G, s_q, e_q)$

  \ForAll{$e \in P$}
    \State $A \leftarrow A \oplus C\big(\textsc{label}(e), e\big)$
  \EndFor

  \State~\Return{$A$}
\end{algorithmic}
\end{algorithm}
}

\newcommand{\deferredCompressionAndSelectFragmentsFigure}{
\begin{figure}[t]
\begin{minipage}{\linewidth}
\begin{minipage}{\columnwidth/2 - 0.25em}
  \centering
  \includegraphics[width=\linewidth, trim=0 0 0 0, clip=true]{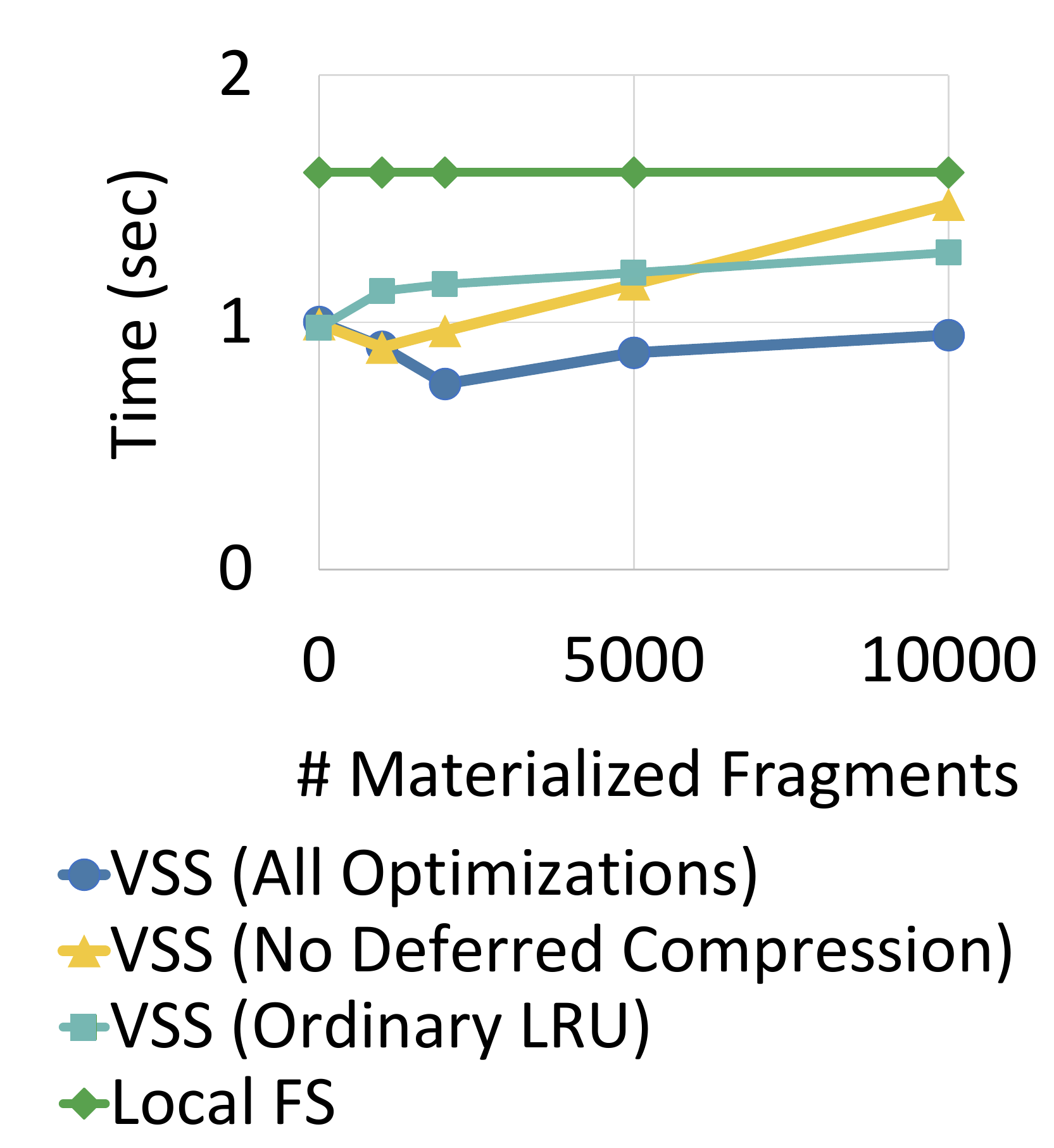}
  \caption{\revision{Selecting and reading short segments.}}
  \label{fig:small-reads}
\end{minipage}\hspace{0.5em}%
\begin{minipage}{\columnwidth/2 - 0.25em}
  \centering
  \includegraphics[width=\linewidth, trim=0 4.25cm 16.25cm 0, clip=true]{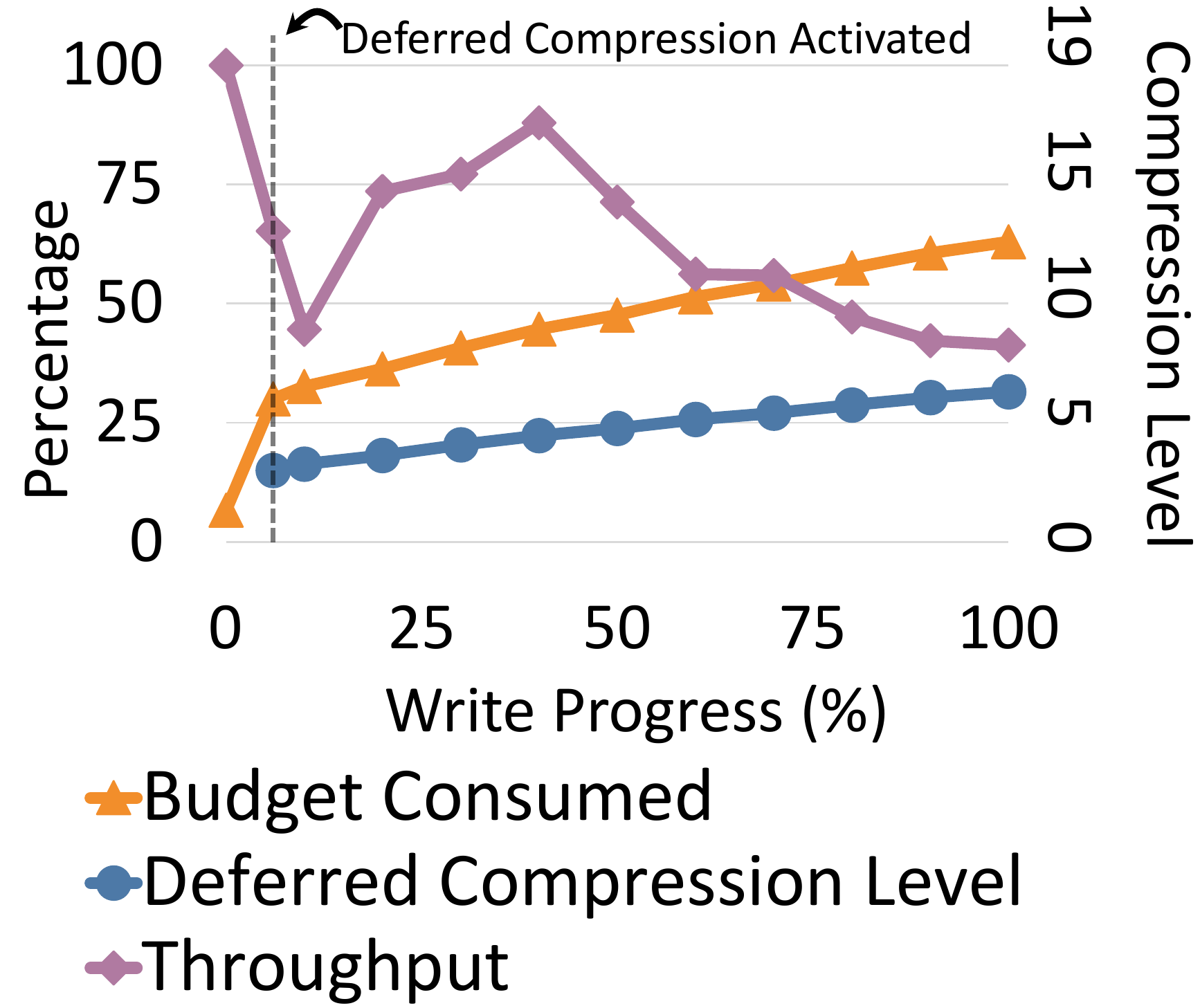}
  \caption{Writes with deferred compression.}
  \label{fig:deferred-compression}
\end{minipage}
\end{minipage}
\end{figure}
}

\newcommand{\applicationAndReadThroughputFigure}{
\begin{figure}[t]
\begin{minipage}{\columnwidth/2 - 0.25em}
  \centering
  \includegraphics[width=\linewidth, trim=0 12em 0 0, clip=true]{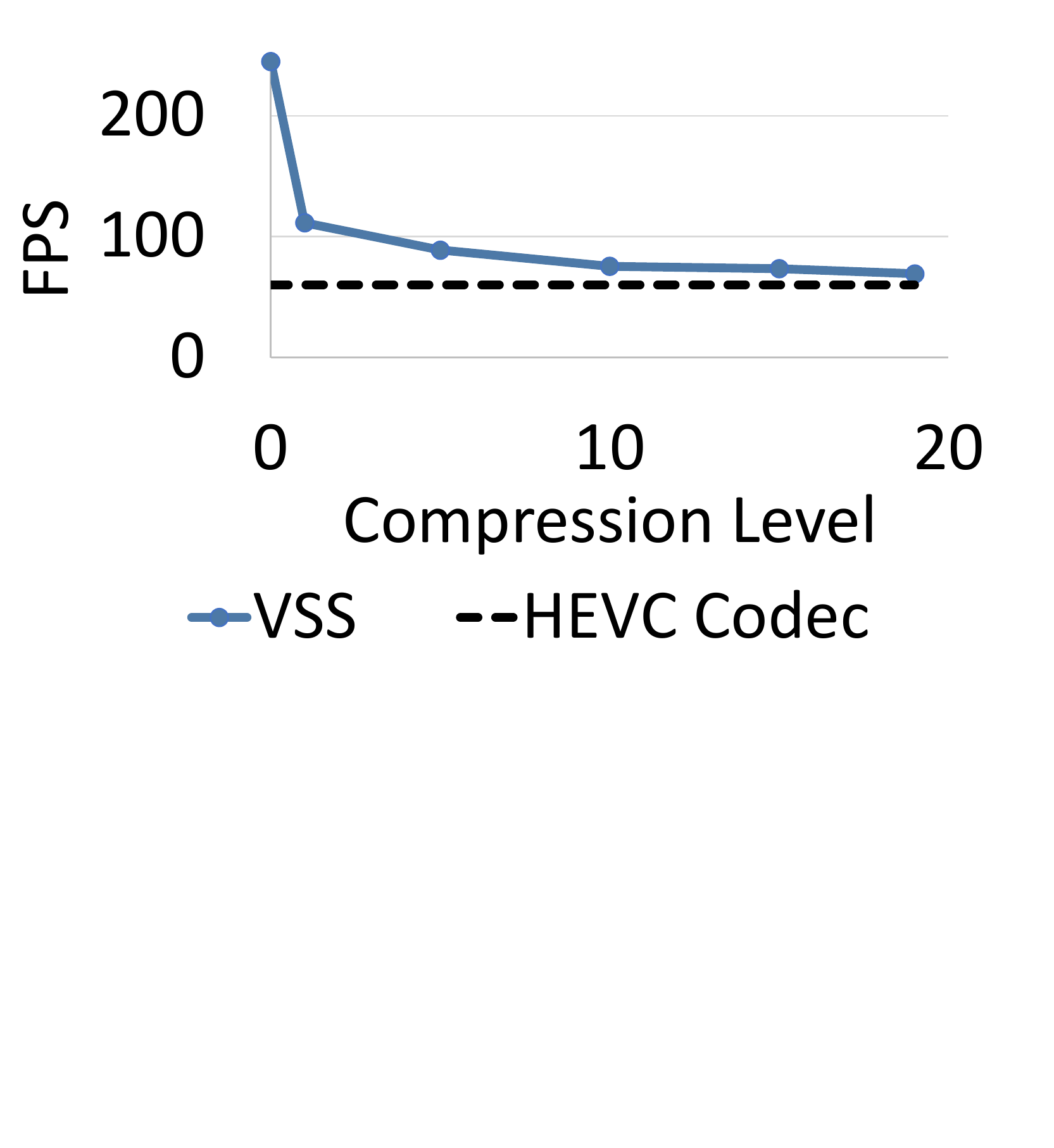}
  \caption{\revision{Throughput for reads over fragments with deferred compression.}}
  \label{fig:deferred-compression-reads}
\end{minipage}\hspace{0.5em}%
\begin{minipage}{\columnwidth/2 - 0.25em}
  \centering
  \includegraphics[width=\linewidth, trim=0 9.5cm 0 0, clip=true]{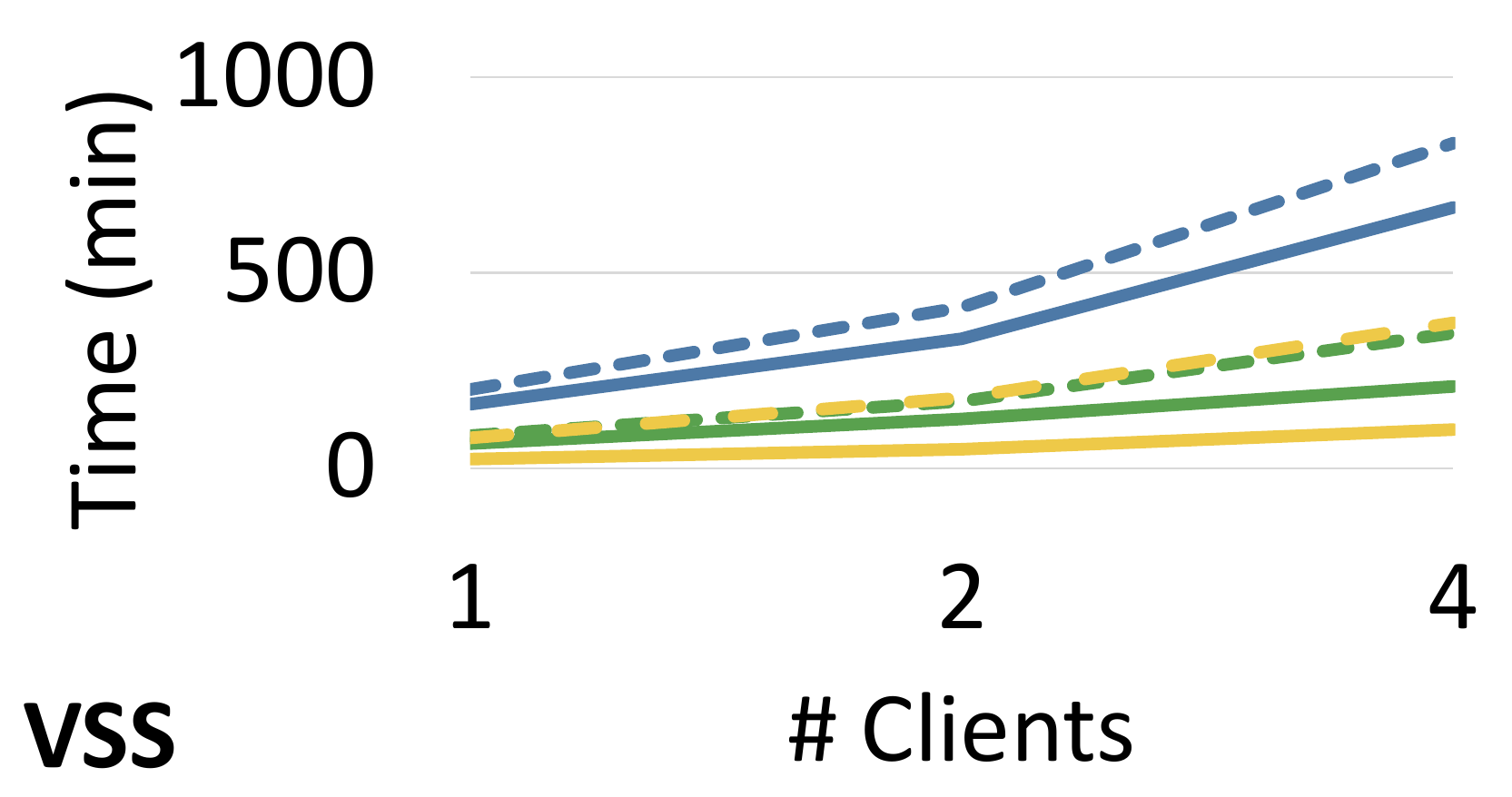}
  \includegraphics[width=\linewidth, trim=0 0 0 14.75cm, clip=true]{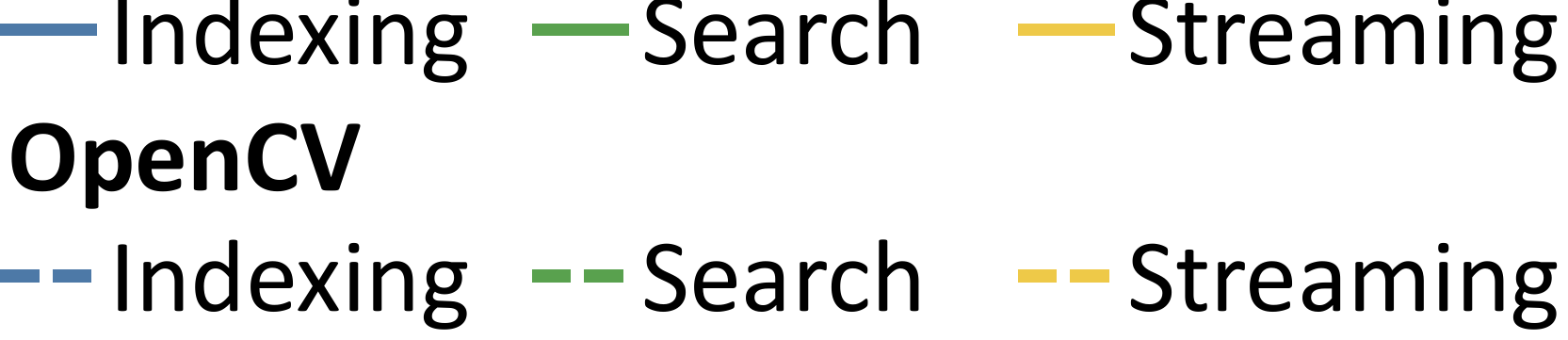}
  \caption{\revision{End-to-end application performance.}}
  \label{figure:end-to-end}
\end{minipage}
\end{figure}
}

\newcommand{\applicationPerformanceFigure}{
\begin{figure}[t]
\centering
\includegraphics[width=\columnwidth/2, trim=0 9.5cm 0 0, clip=true]{figures/analytical-tasks.pdf}
\includegraphics[width=\columnwidth/2, trim=0 0 0 14.75cm, clip=true]{figures/analytical-tasks-legend.pdf}
\caption{\revision{End-to-end application performance of \name and an OpenCV variant.  The application is decomposed into indexing, search, and streaming subtasks and executed by various batch sizes.}}
\label{figure:end-to-end}
\end{figure}
}

\newcommand{\vssCompression}{
\begin{figure}[t]
\centering
\includegraphics[width=0.95\columnwidth]{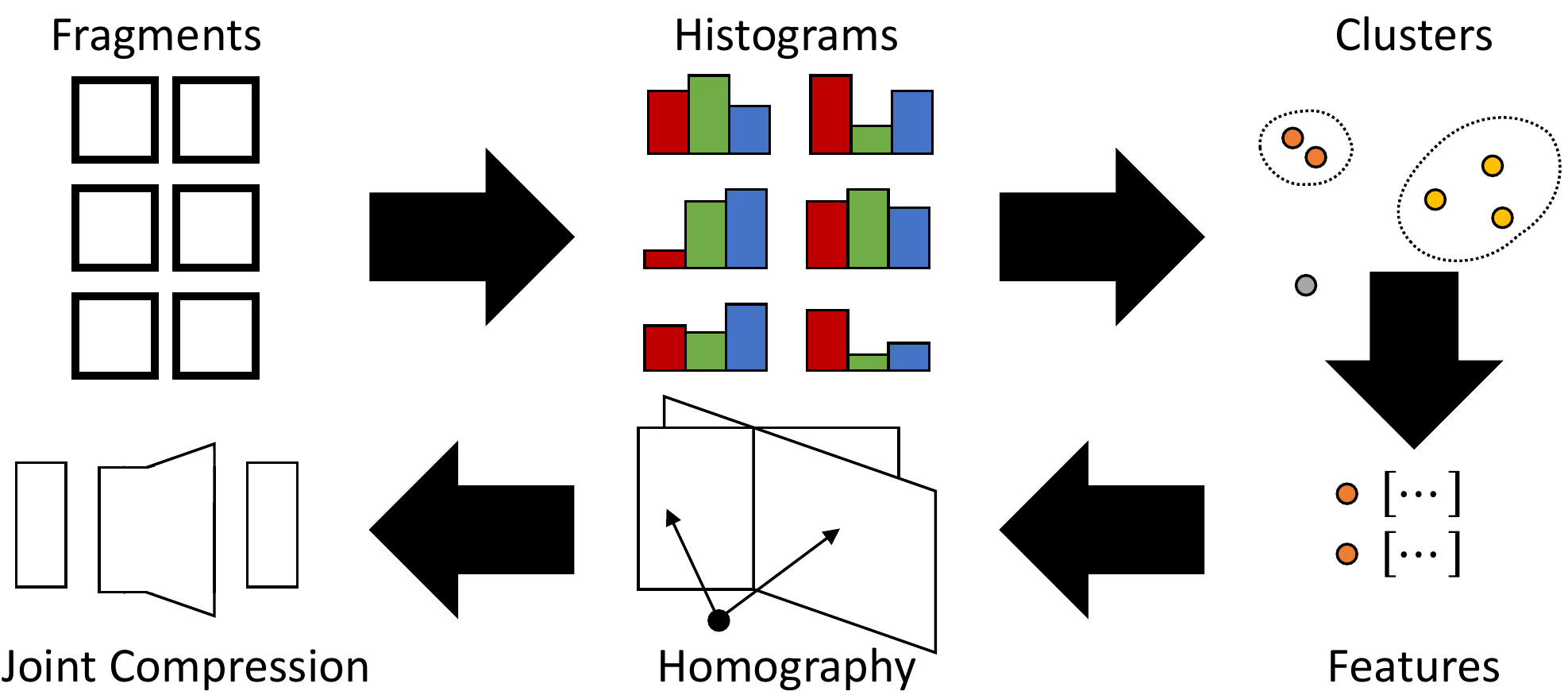} 
\caption{\squeezemore{Joint compression fragment selection process: (i) compute and cluster fragment histograms, (ii) for the smallest cluster, compute features and search for fragments with many similar features, and (iii) for each pair, compute homography and (iv) compress.}}
\label{figure:compression-overview}
\end{figure}
}

\newcommand{\vssPhysicalOrganization}{
\begin{figure}[t]
\centering
\includegraphics[width=0.75\columnwidth]{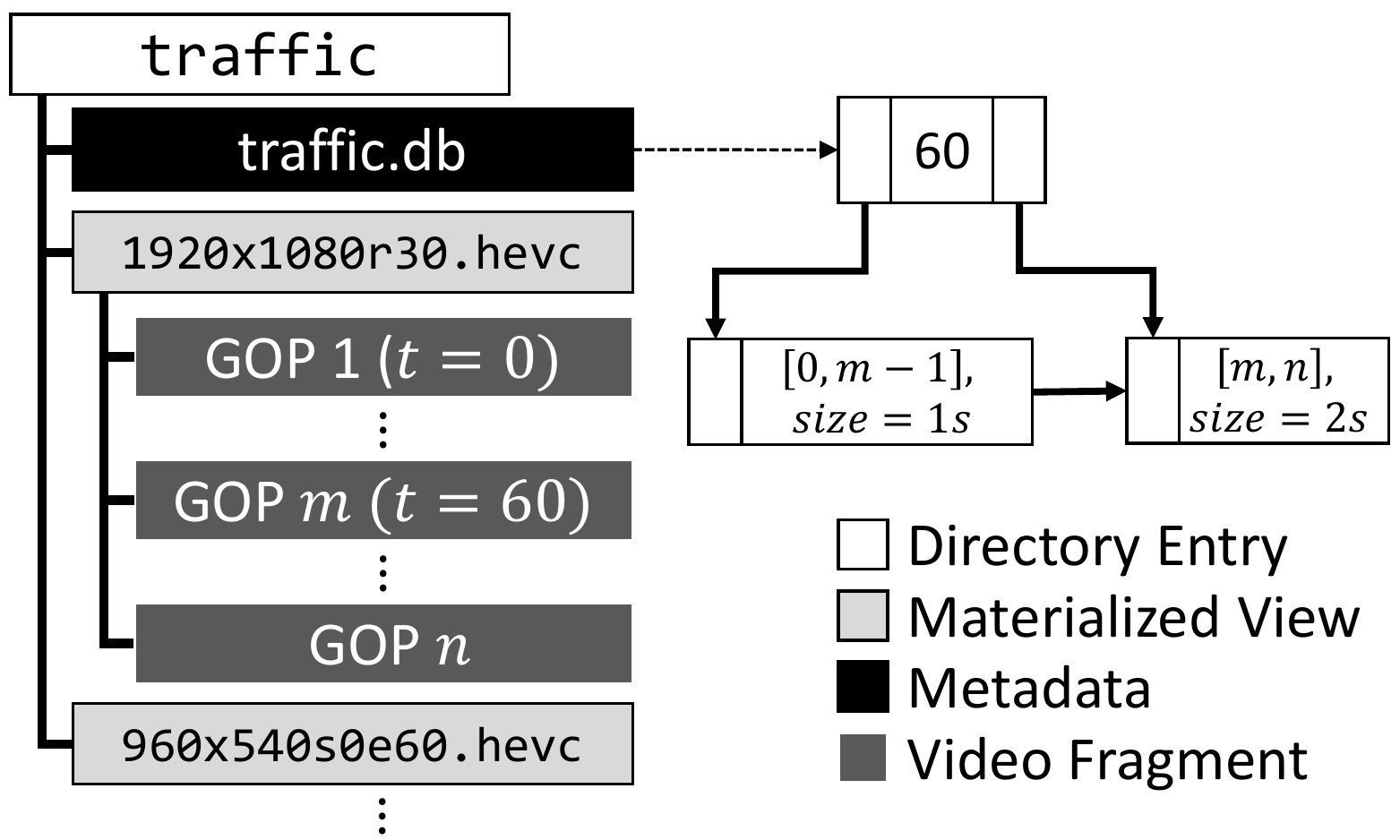} 
\caption{\squeeze{An example \name physical organization that contains one logical video and two underlying physical videos.  For physical video \texttt{1920x1080r30.hevc}, the first $m$ GOPs are each one second in length, while the remaining $n-m$ are two seconds.  These durations are recorded in the associated index.}}
\label{figure:vss-physical}
\end{figure}
}

\newcommand{\vssCaching}{
\begin{figure}[t]
\centering
\includegraphics[width=\columnwidth]{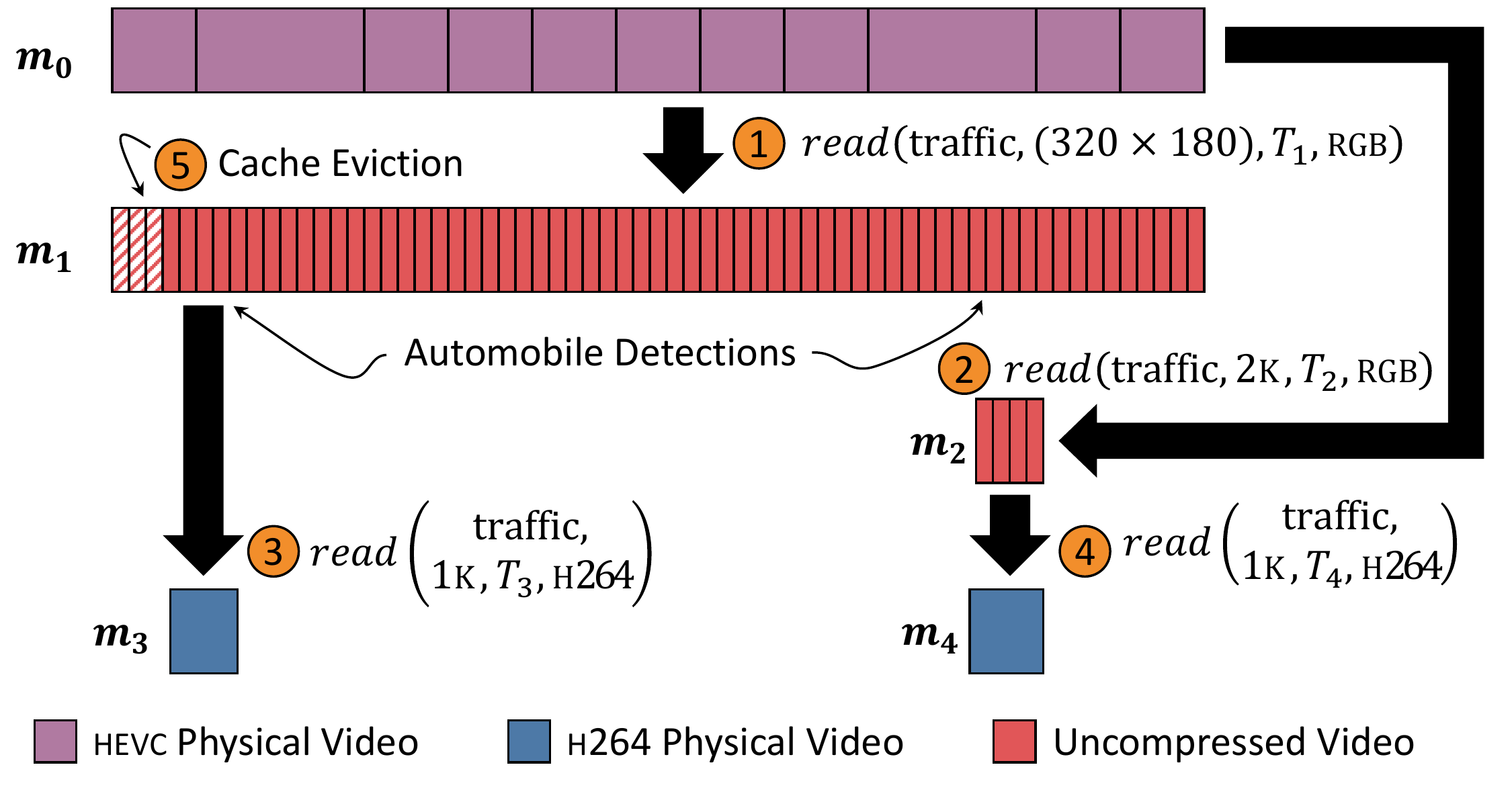} 
\caption{\squeezemore{\name caches read results and uses them to answer future queries.  In \protect\circled{1} an application reads \texttt{traffic} at $320{\times}180$ resolution for use in object detection, which \name caches as $m_1$.  In \protect\circled{2} \name caches $m_2$, a region with a dubious detection.  In \protect\circled{3} and \protect\circled{4} \name caches \avc-encoded $m_3$ \& $m_4$, where objects were detected.  However, reading $m_4$ exceeds the storage budget and \name evicts the striped region at \protect\circled{5}.}}
\label{figure:caching}
\end{figure}
}

\newcommand{\lookbackCostFigure}{
\begin{figure}[t]
\centering
\includegraphics[width=0.95\columnwidth]{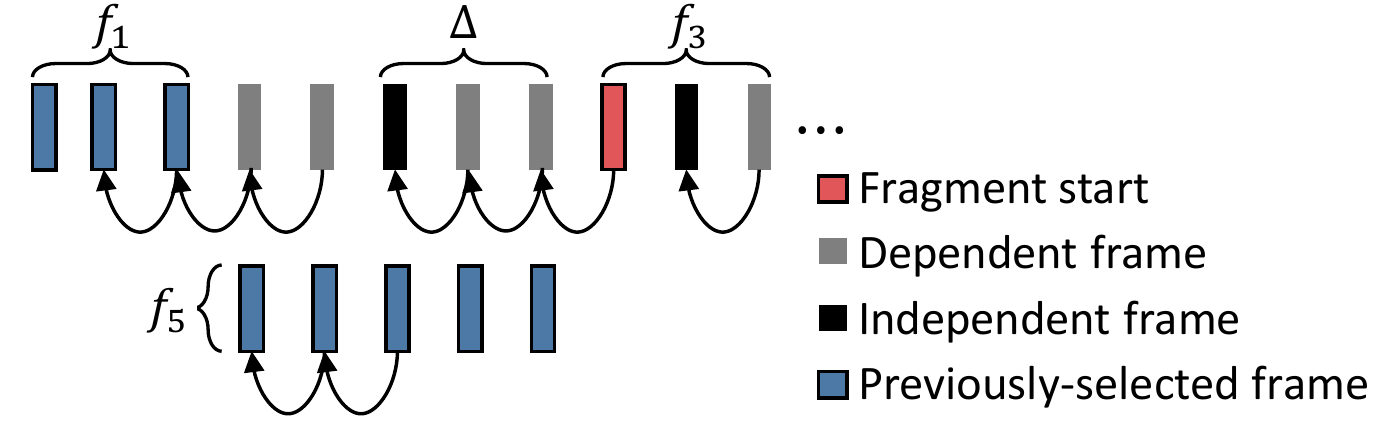} 
\caption{\squeeze{A simplified illustration based on \autoref{fig:query-answering}.  \name has decided to use $f_1$ and $f_5$ and is considering using $f_3$ starting at the red-highlighted frame.  However, $f_3$ cannot be decoded without transitively decoding its dependencies shown by directed edges (labeled $\Delta$).}}
\label{figure:lookback-cost}
\end{figure}
}

\newcommand{\queryAnsweringFigure}{
\begin{figure}[t]
\centering
\subfigure[\columnwidth][Read operation on three materialized physical videos]{
  \includegraphics[width=\columnwidth, trim=0 0.5cm 0 0, clip=true]{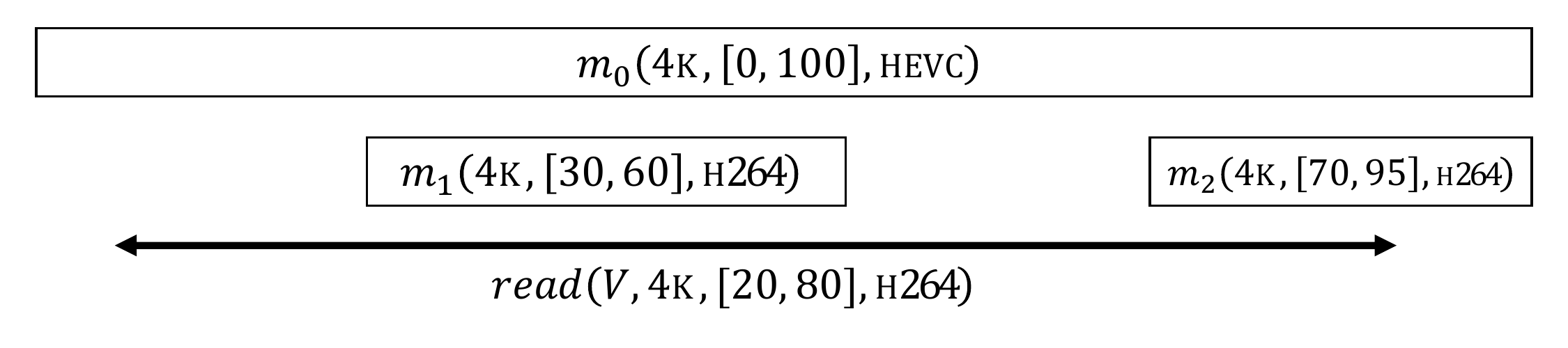} 
\label{subfig:query-answering-views}
}\\
\subfigure[\columnwidth][Physical video fragments with simplified cost formulae]{
  \includegraphics[width=\columnwidth]{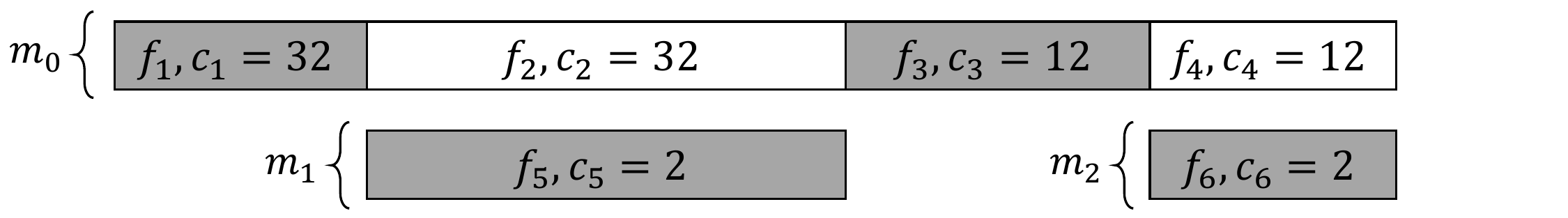} 
\label{subfig:query-answering-fragments}
}\\
\caption{\squeezemore{\Cref{subfig:query-answering-views} shows the query $read(V, 4\normalfont{\textsc{k}},$ $[20, 80], \avc)$, where \name has materialized $m_0$, $m_1$, and $m_2$.}  \Cref{subfig:query-answering-fragments} shows weighted fragments and costs.  The lowest-cost result is shaded.}
\label{fig:query-answering}
\end{figure}
}

\newcommand{\projectionFigure}{
\begin{figure}[t]
  \centering
  \subfigure[\columnwidth/2][Frame from left video\hspace{3em}]{
    \hspace{0.5\columnwidth}
    \label{subfig:projection-frame-1}
  }%
  \subfigure[\columnwidth/2][Frame from right video\hspace{1em}]{
    \hspace{0.5\columnwidth}
    \label{subfig:projection-frame-2}
  }\\
  \includegraphics[width=\columnwidth/2]{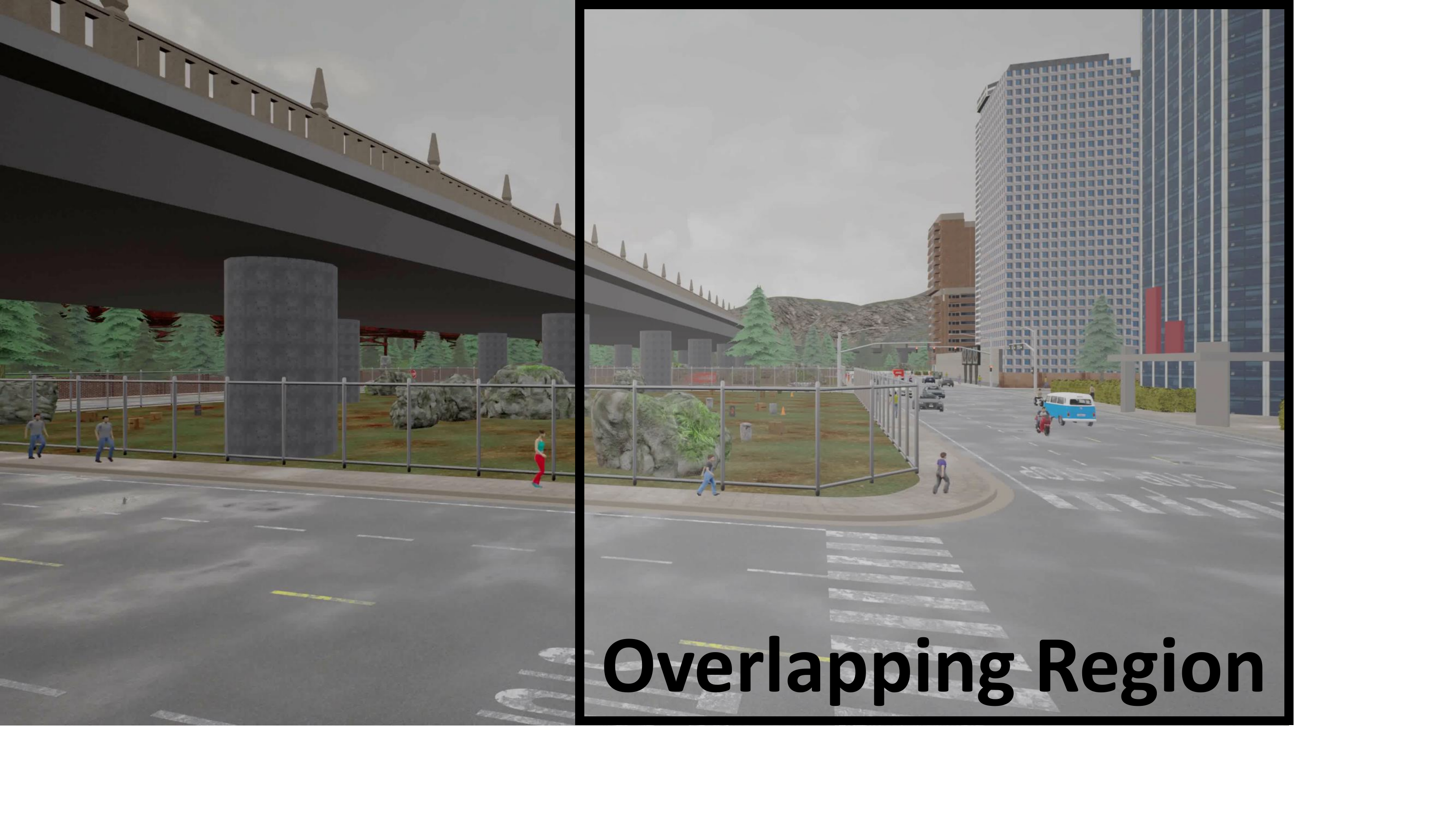}%
  \includegraphics[width=\columnwidth/2]{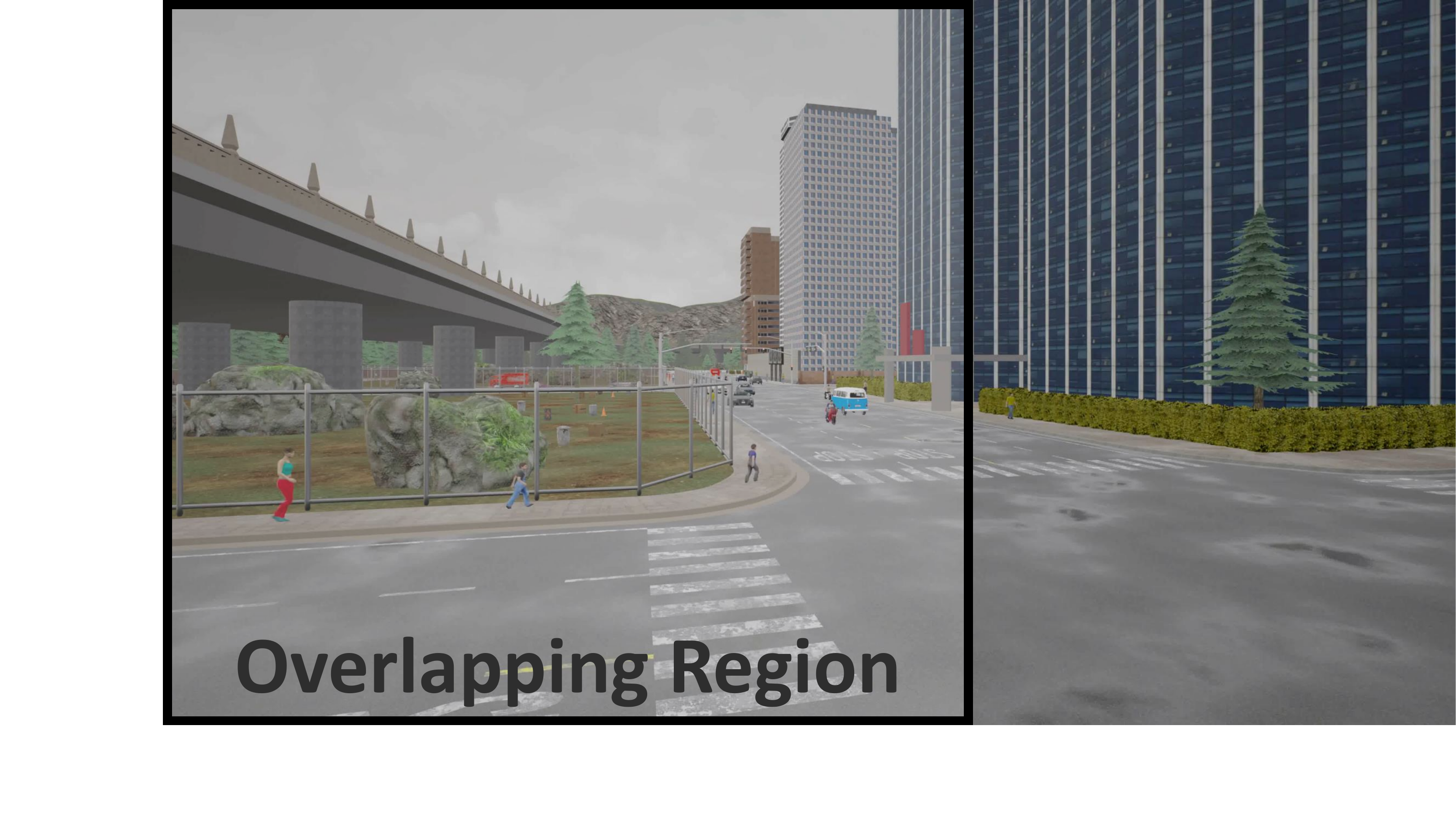}%
  \\
  \subfigure[\textwidth/3][Left, overlapped, and right regions are separately encoded.]{
    \includegraphics[width=\columnwidth]{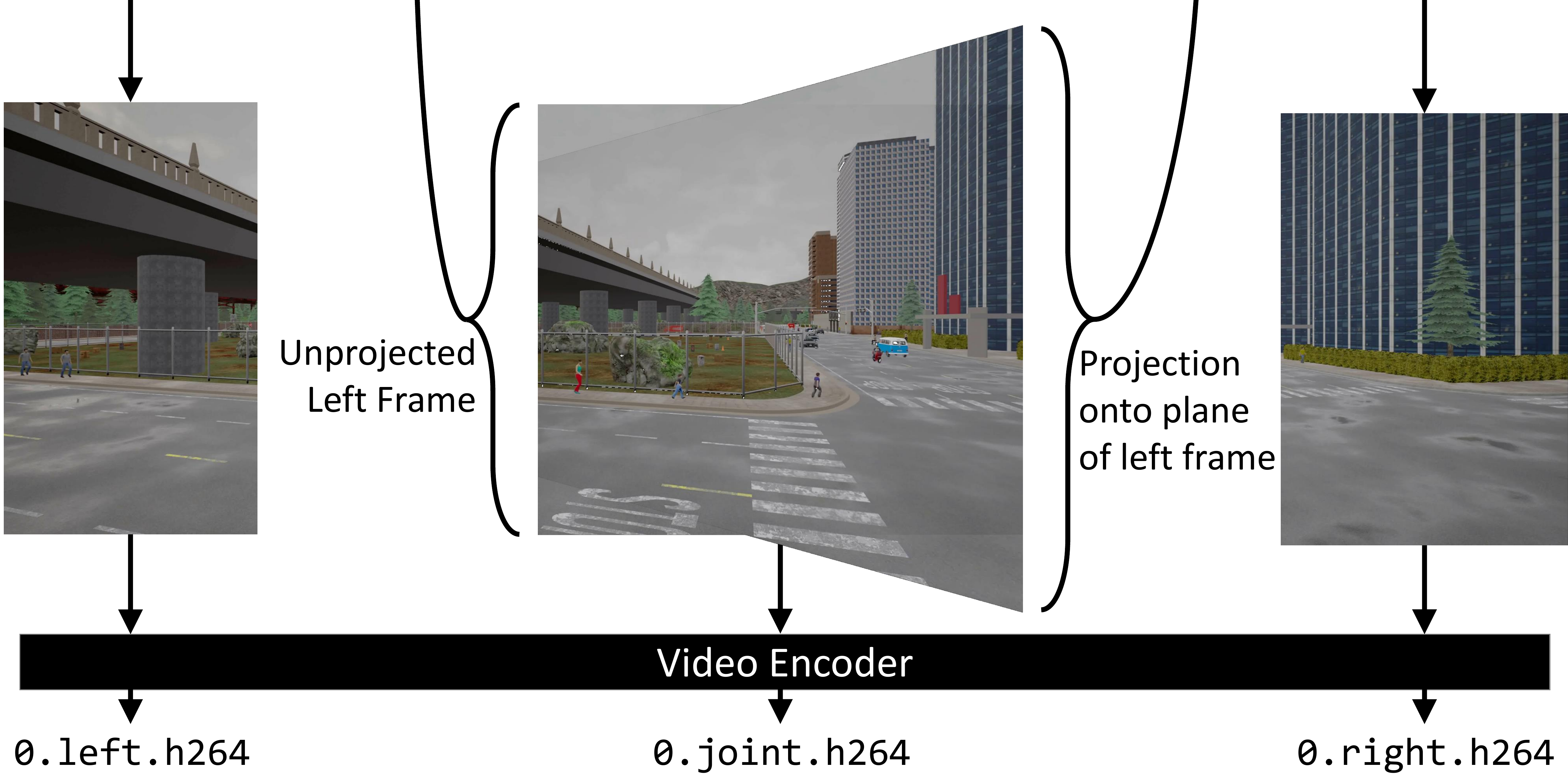}
  \label{subfig:projection-overlap}
  }
  \caption{\squeezemore{The joint compression process.  \name identifies overlap, combines it, and separately encodes the pieces.}}
  \label{fig:projection}
\end{figure}
}

\newcommand{\readFigure}{
\begin{figure}[t]
  \begin{minipage}{\linewidth}
  \centering
  \includegraphics[width=\columnwidth, trim=0 0 0 9cm, clip=true]{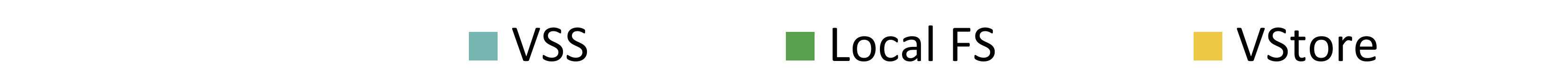}
  \end{minipage}
  \subfigure[\columnwidth/2][Read in same format]{
    \includegraphics[width=\columnwidth/2, trim=0 2cm 16cm 0, clip=true]{figures/read-6-1-same-format.pdf}
    \label{subfig:read-same-format}
  }%
  \subfigure[\columnwidth/2][Read in different format]{
    \includegraphics[width=\columnwidth/2, trim=16cm 2cm 0 0, clip=true]{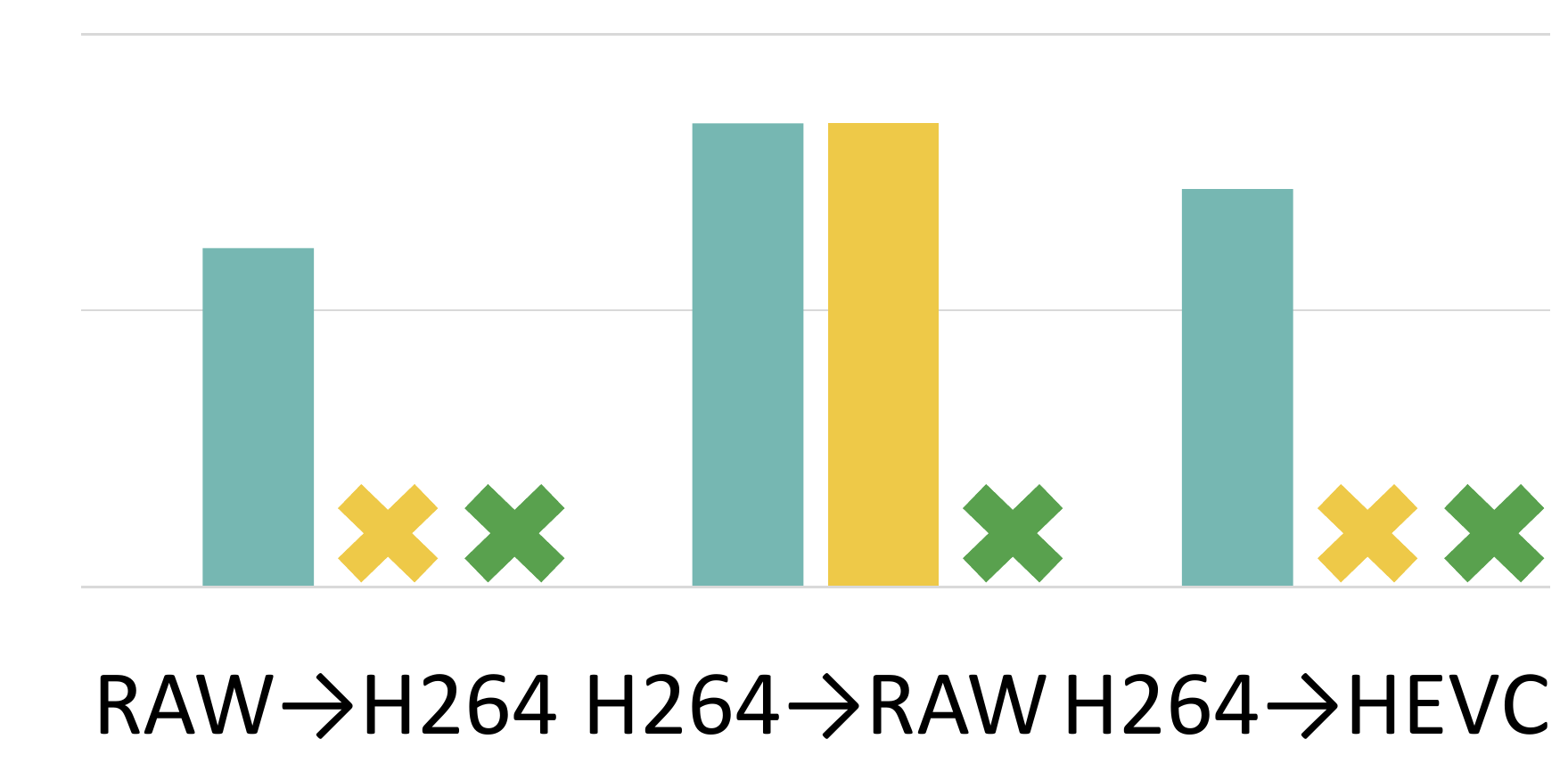}
    \label{subfig:read-different-format}
  }%
  \caption{Read throughput.  Each group $I{\rightarrow} O$ shows throughput reading in format $I$ and outputting in format $O$.  
  \squeezemore{An $\boldsymbol{\times}$ indicates lack of support for a read type.}}
  \label{fig:read}
\end{figure}
}

\newcommand{\projectionResultsFigure}{
\begin{figure}[t]
  \centering
  \subfigure[\columnwidth/2][Recovered frame from left video]{
    \includegraphics[width=\columnwidth/2]{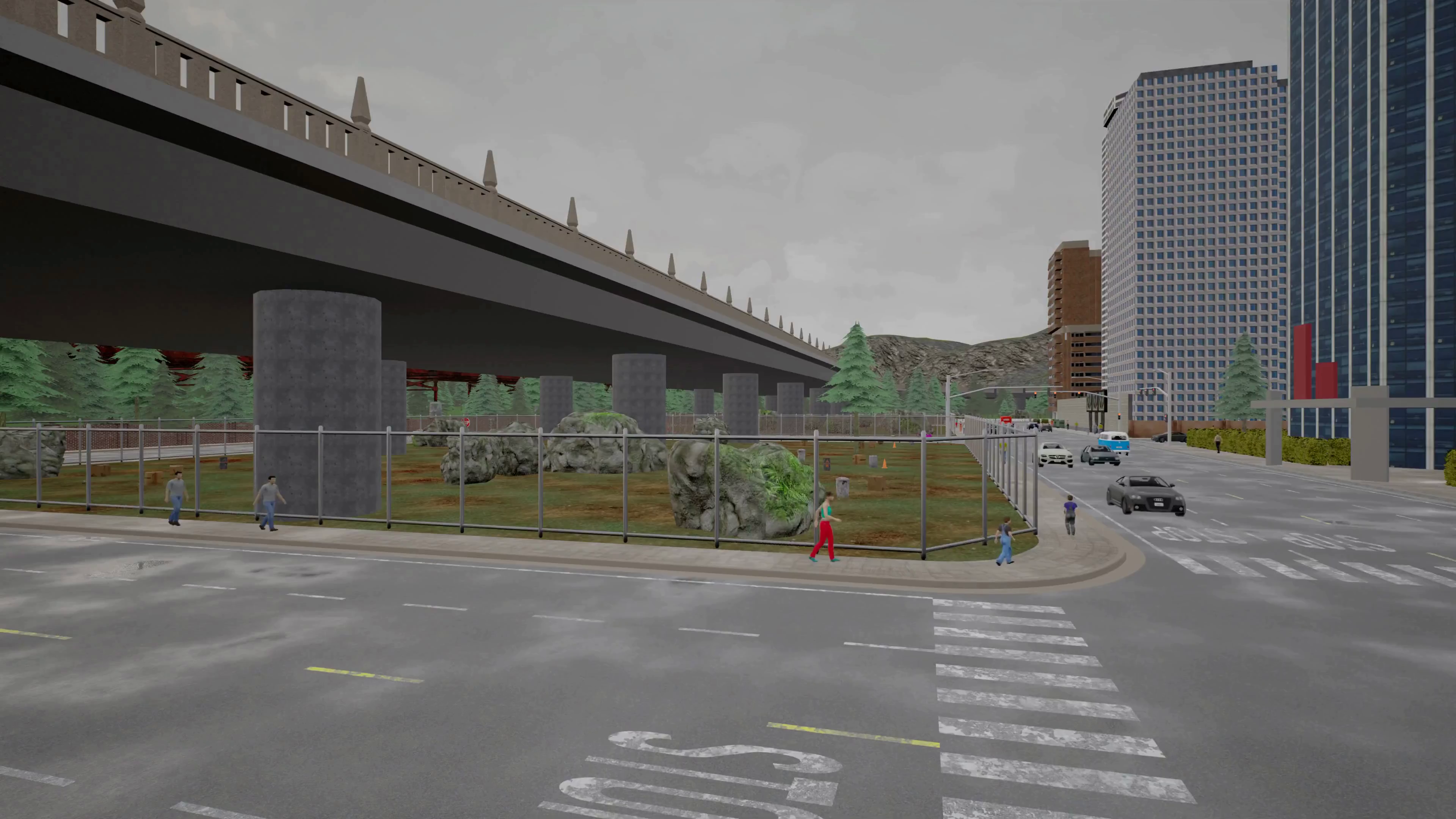}
  \label{subfig:recovered-left-leftjoin}
  }%
  \subfigure[\columnwidth/2][Recovered frame from right video]{
    \includegraphics[width=\columnwidth/2]{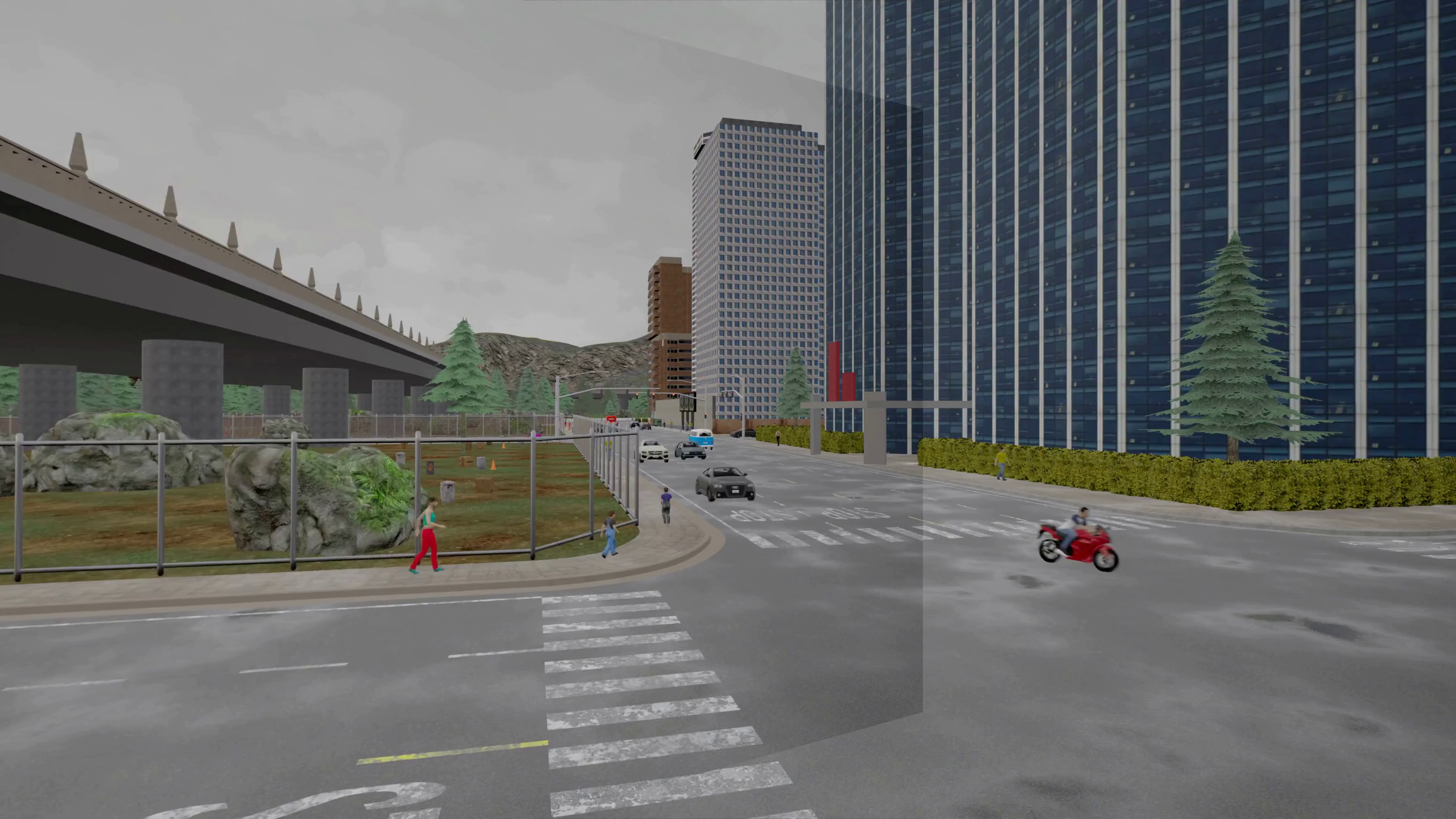}
  \label{subfig:recovered-right-leftjoin}
  }%
  \caption{Recovered frames from joint compression.}
  \label{fig:recovered-leftjoin}
\end{figure}
\begin{figure}[t]
  \centering
  \subfigure[\columnwidth/2][Original frame]{
    \includegraphics[width=\columnwidth/2]{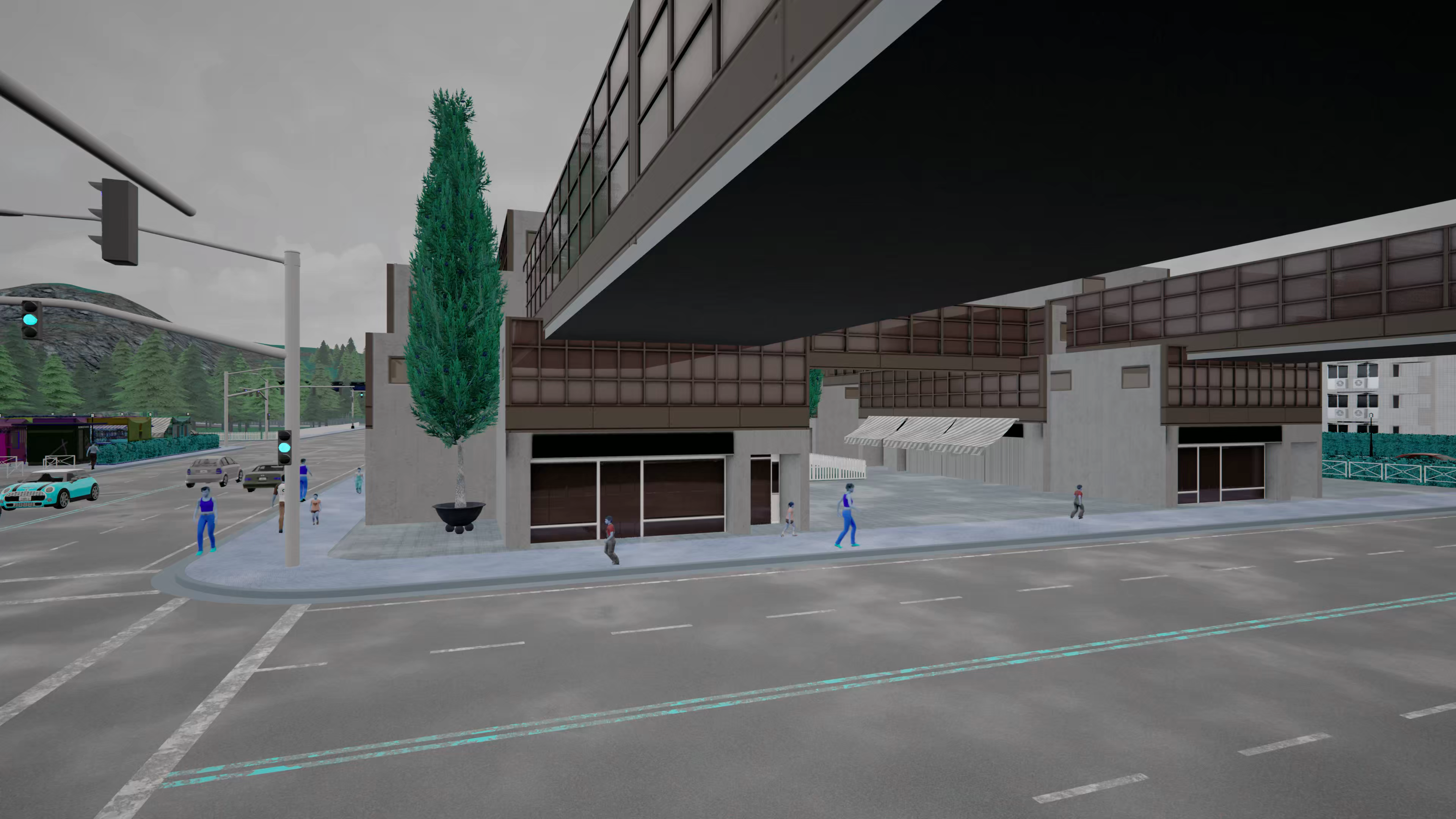}
  \label{subfig:aborted-left-leftjoin}
  }%
  \subfigure[\columnwidth/2][Recovered frame]{
    \includegraphics[width=\columnwidth/2]{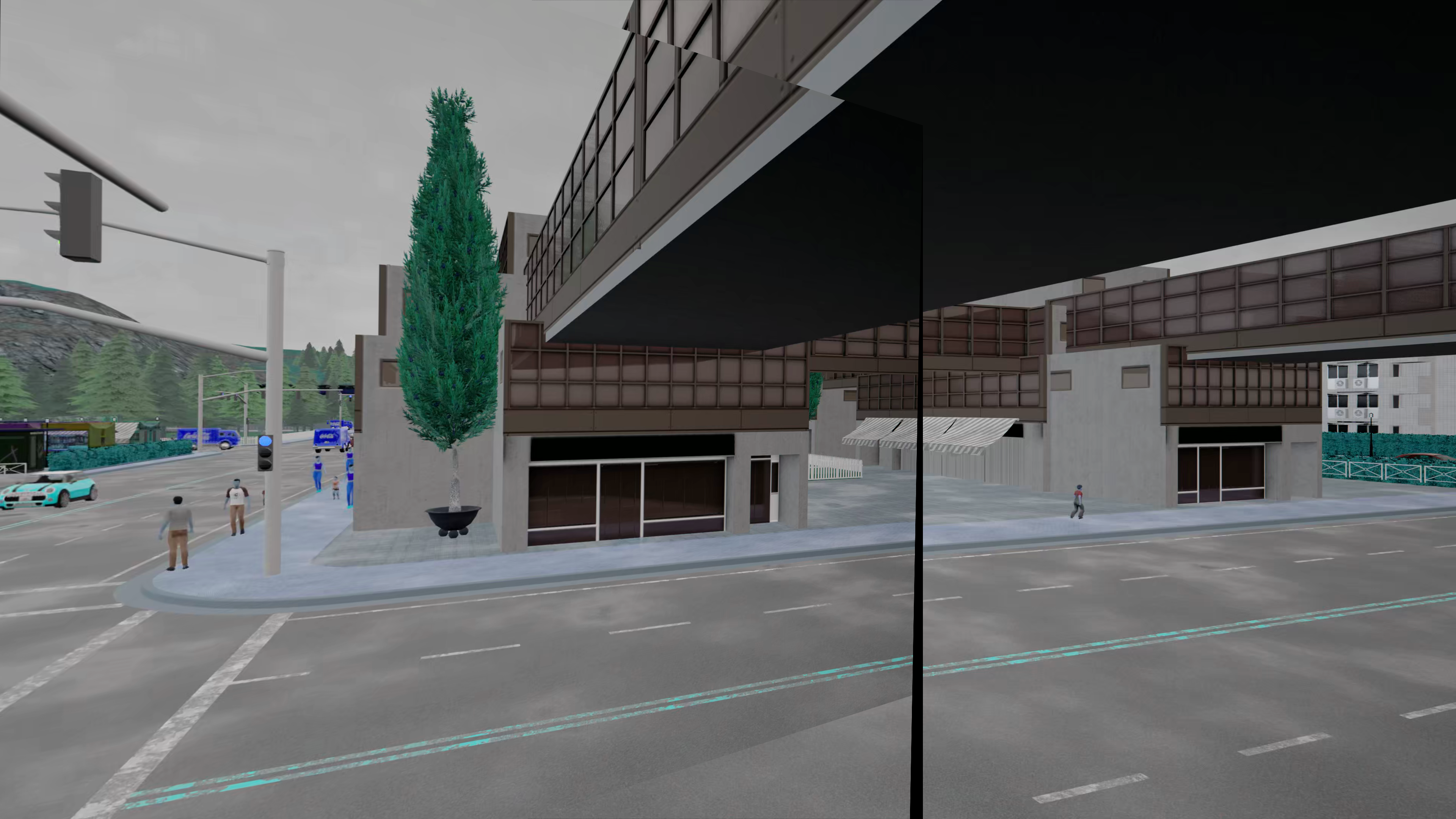}
  \label{subfig:aborted-right-leftjoin}
  }%
  \caption{An example recovered frame with slightly incorrect homography.  In this case \name detects the error relative to the original frame and aborts joint compression.}
  \label{fig:aborted-leftjoin}
\end{figure}
}

\newcommand{\fragmentReadFigure}{
\begin{figure}[t]
\begin{minipage}{\linewidth}
\begin{minipage}{\columnwidth/2 - 0.25em}
  \centering
  \includegraphics[width=\linewidth, trim=0 3cm 0 0, clip=true]{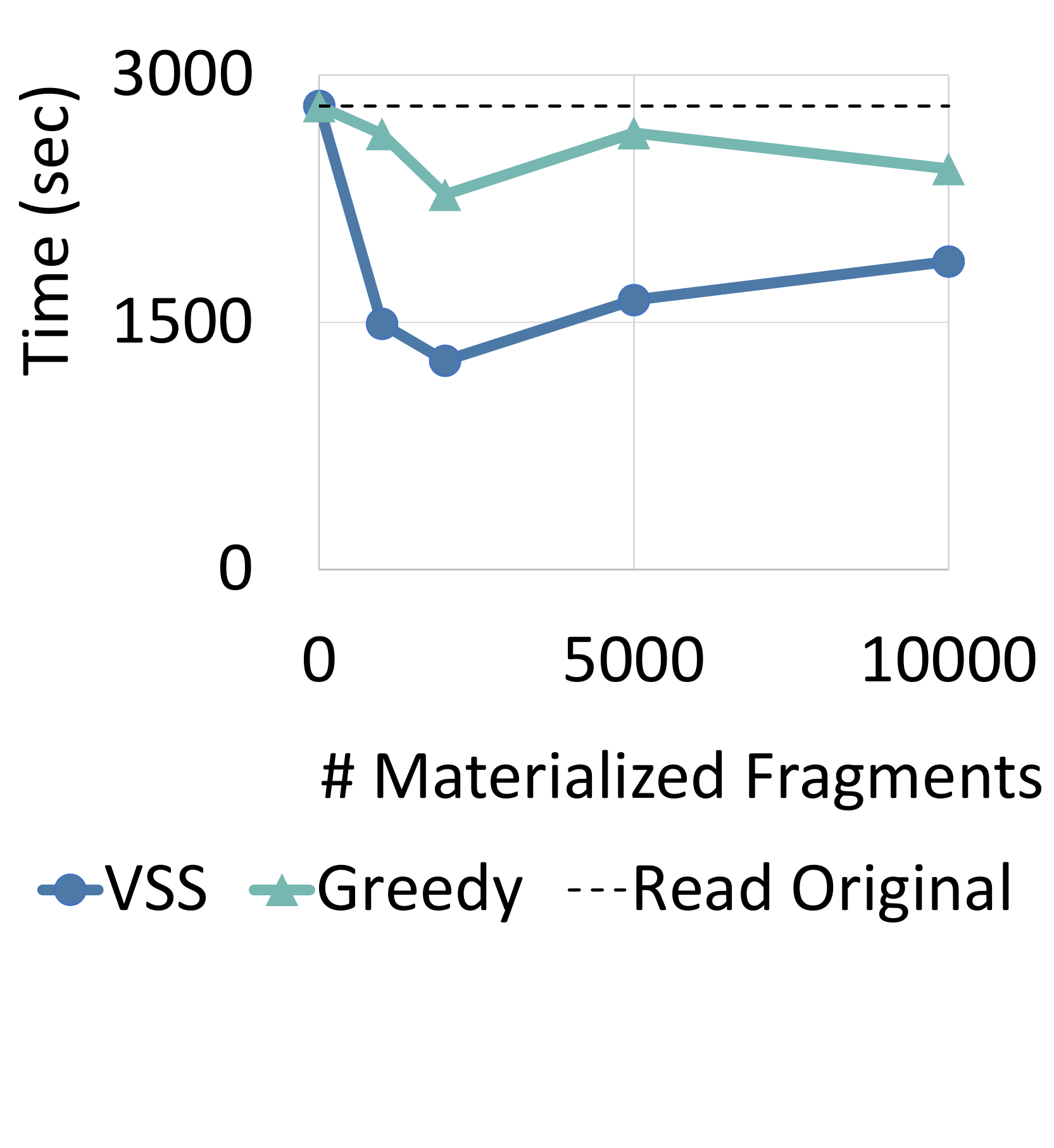}
  \caption{Time to select fragments and read video.}
  \label{fig:read-fragments}
\end{minipage}\hspace{0.5em}%
\begin{minipage}{\columnwidth/2 - 0.25em}
  \centering
  \includegraphics[width=\linewidth, trim=0 3cm 0 0, clip=true]{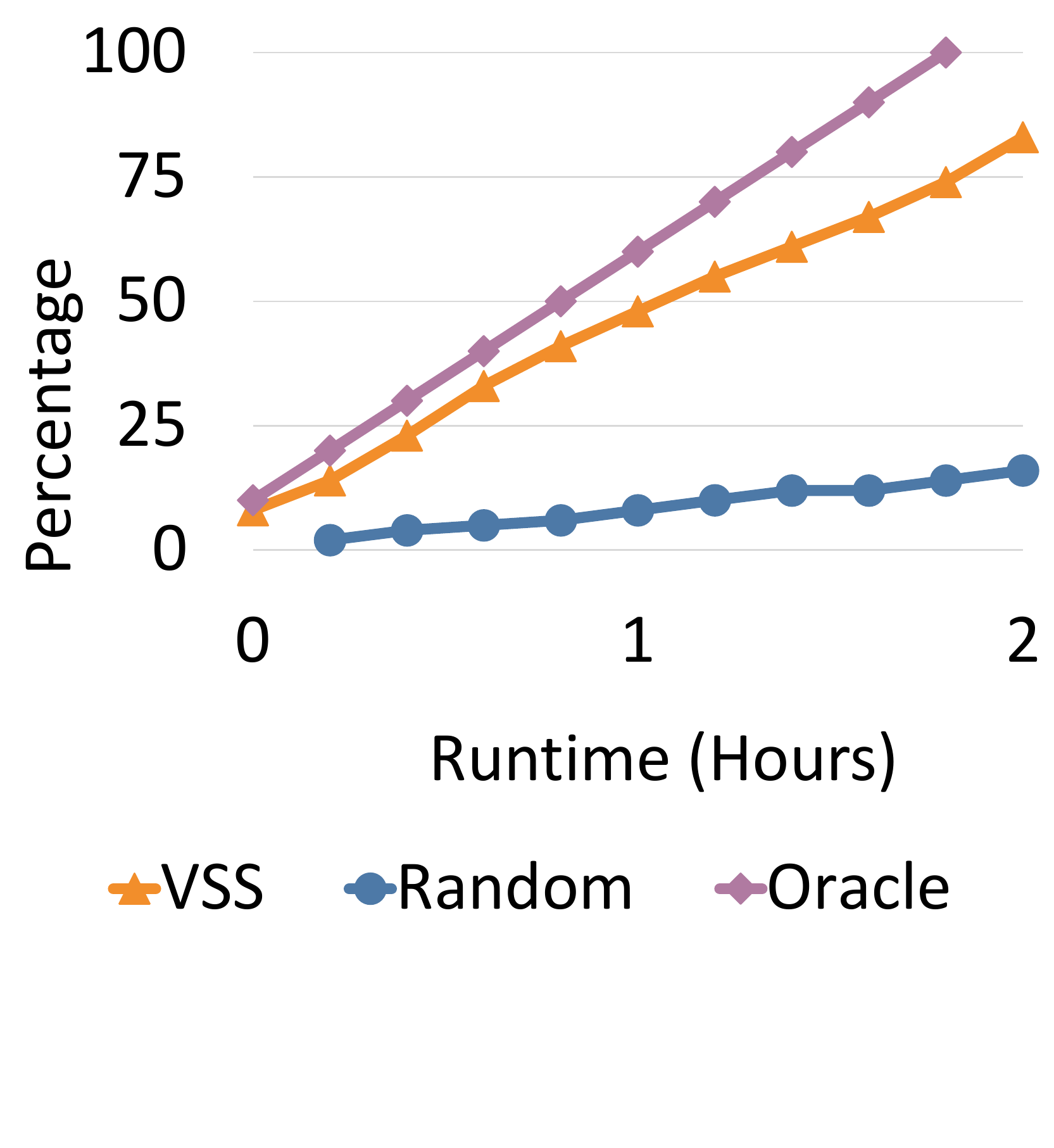}
  \caption{\textls[-55]{Joint compression pair selection.}}
  \label{fig:compression-selection-method}
\end{minipage}
\end{minipage}
\end{figure}
}

\newcommand{\deferredCompressionReadFigure}{
\begin{figure}[t]
\centering
\includegraphics[width=0.5\linewidth, trim=0 12em 0 0, clip=true]{figures/deferred-compression-reads.pdf}
\caption{\revision{Throughput for reads over fragments with deferred compression applied at various levels.}}
\label{fig:deferred-compression-reads}
\end{figure}
}

\newcommand{\jointCompressionThroughputFigure}{
\begin{figure}[t]
  \centering
  \begin{minipage}{\linewidth}
  \centering
  \includegraphics[width=\columnwidth, trim=0 1.5em 0 16.75cm, clip=true]{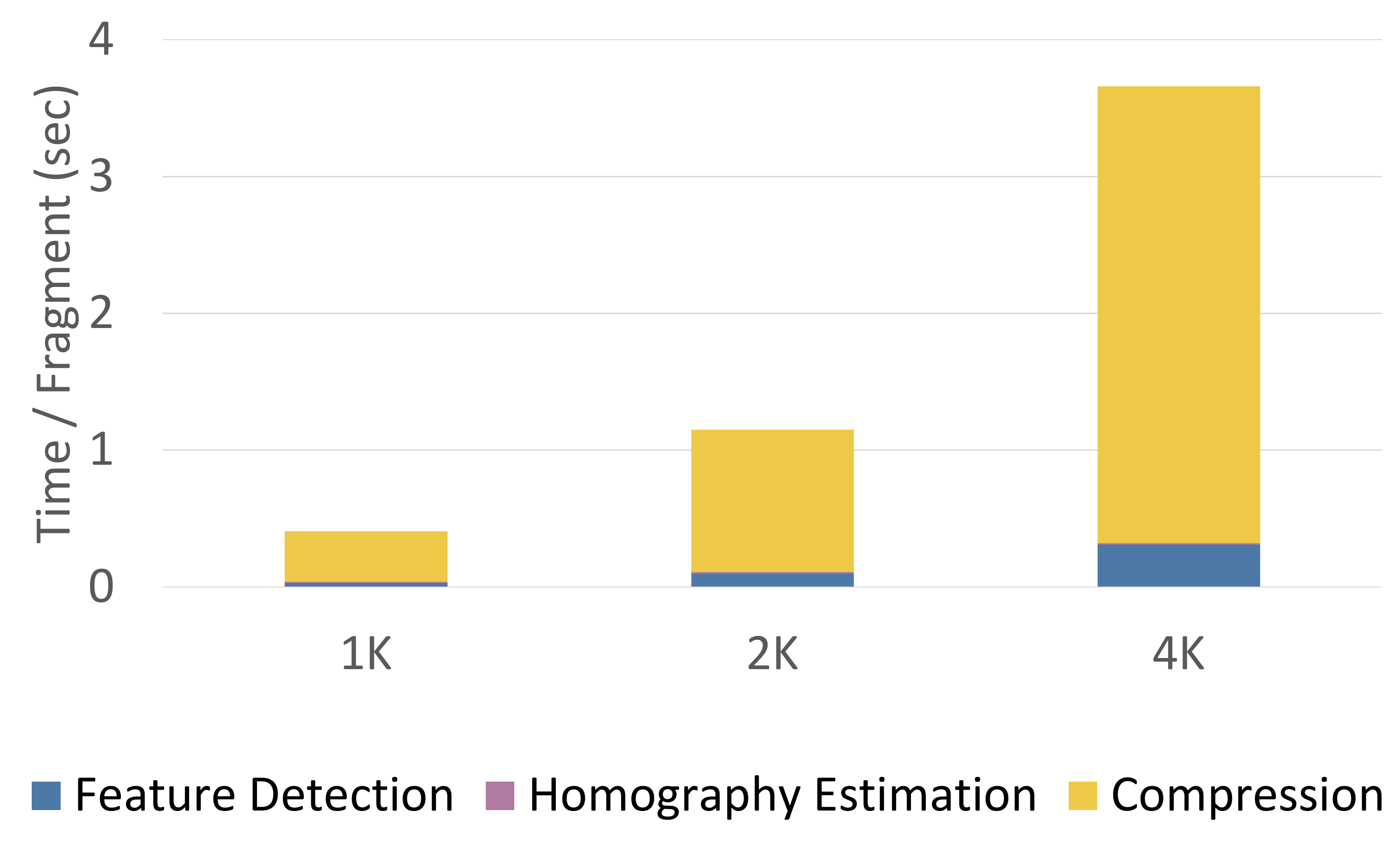}
  \end{minipage}
  \subfigure[\columnwidth/2][By resolution]{
    \includegraphics[width=0.9\columnwidth/2, trim=0 27em 0 0, clip=true]{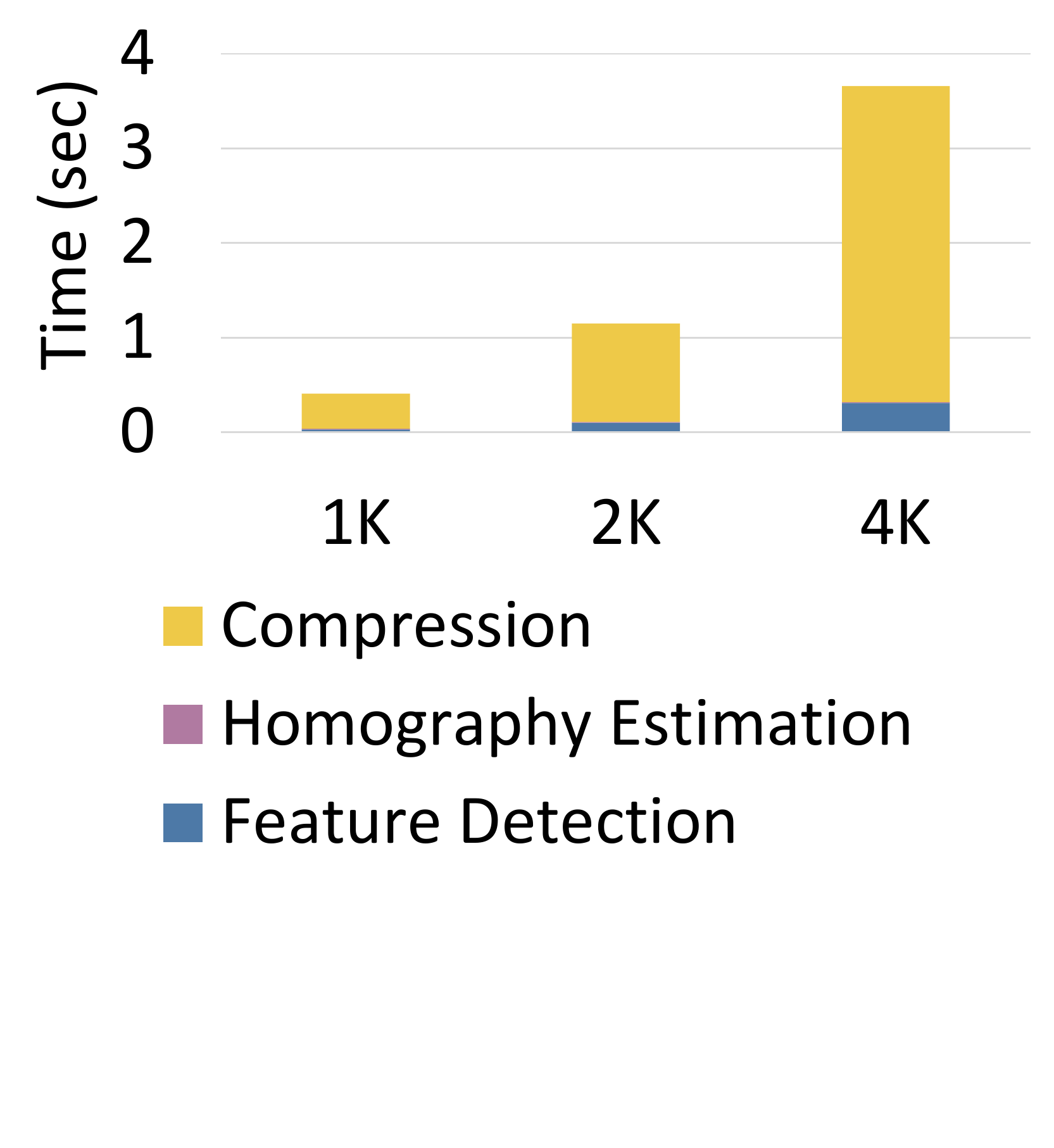}
    \label{subfig:joint-compression:resolution}
  }%
  \subfigure[\columnwidth/2][By dynamicism]{
    \includegraphics[width=0.9\columnwidth/2, trim=0 27em 0 0, clip=true]{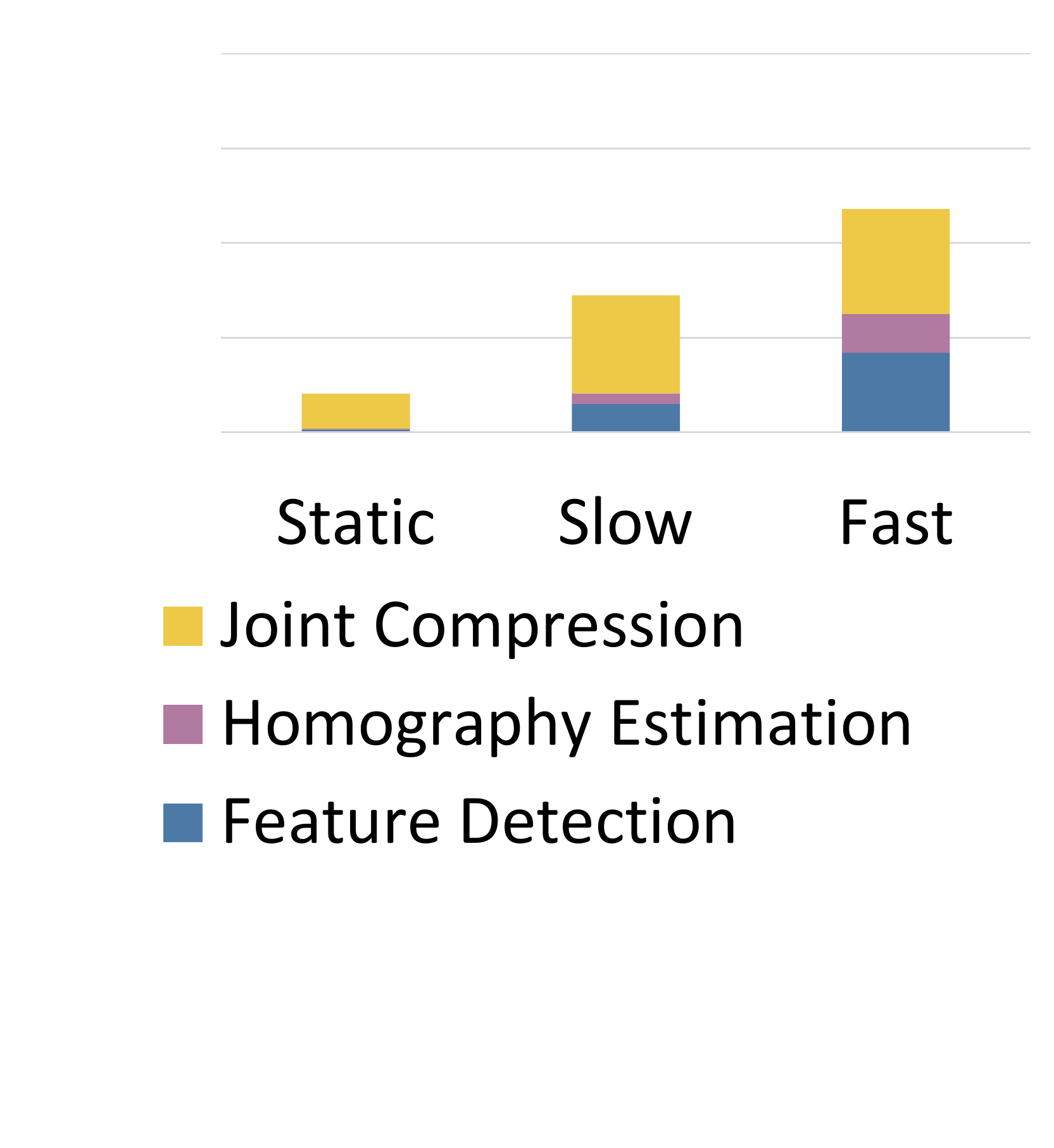}
    \label{subfig:joint-compression:dynamic}
  }
  \caption{\revision{Joint compression overhead.}}
  \label{fig:joint-compression-throughput}
\end{figure}
}

\newcommand{\writeFigure}{
\begin{figure}[t]
  \centering
  \includegraphics[width=\columnwidth, trim=0 0 0 10cm, clip=true]{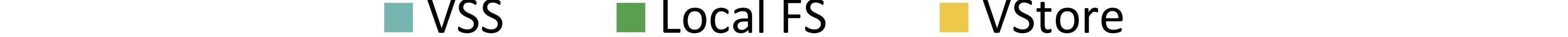}
  \subfigure[\columnwidth][Uncompressed writes]{
    \includegraphics[width=0.95\columnwidth, trim=0 2.25cm 0 0, clip=true]{figures/write.pdf}
    \label{subfig:write:uncompressed}
  }\\
  \subfigure[\columnwidth][Compressed writes (\avc)]{
    \includegraphics[width=0.95\columnwidth, trim=0 2.25cm 0 0, clip=true]{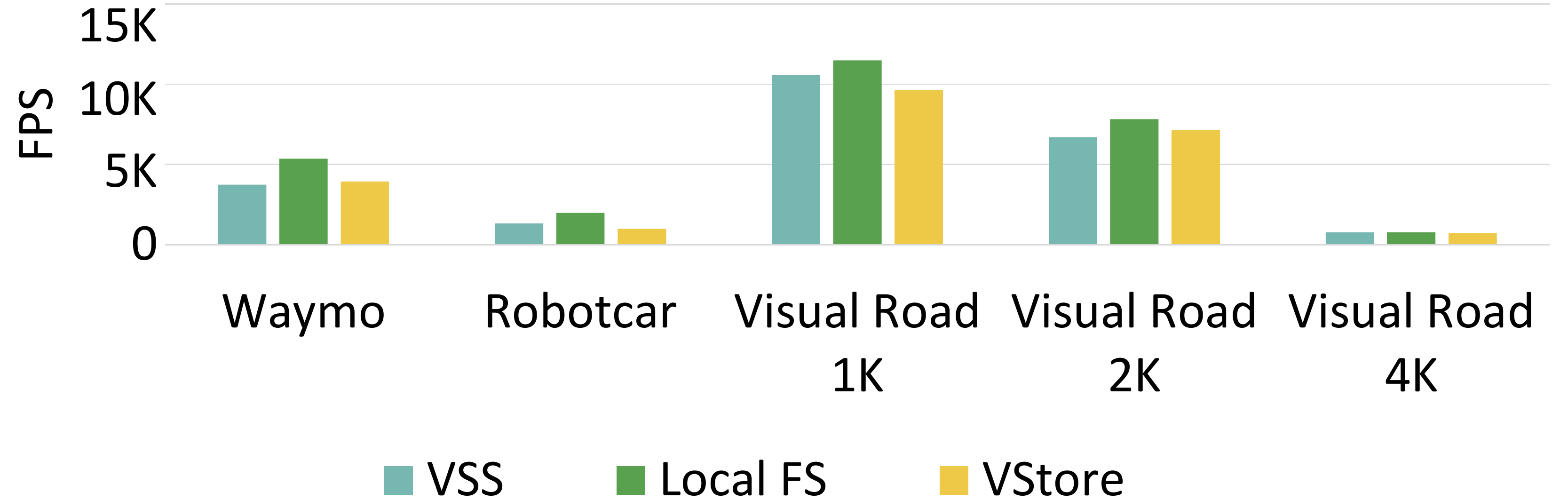}
    \label{subfig:write:compressed}
  }
  \caption{Throughput to write video.}
  \label{fig:write}
\end{figure}
}

\newcommand{\jointThroughputFigure}{
\begin{figure}[t]
\begin{minipage}{\linewidth}
\centering
\includegraphics[width=0.75\columnwidth, trim=0 0.25em 0 9.6cm, clip=true]{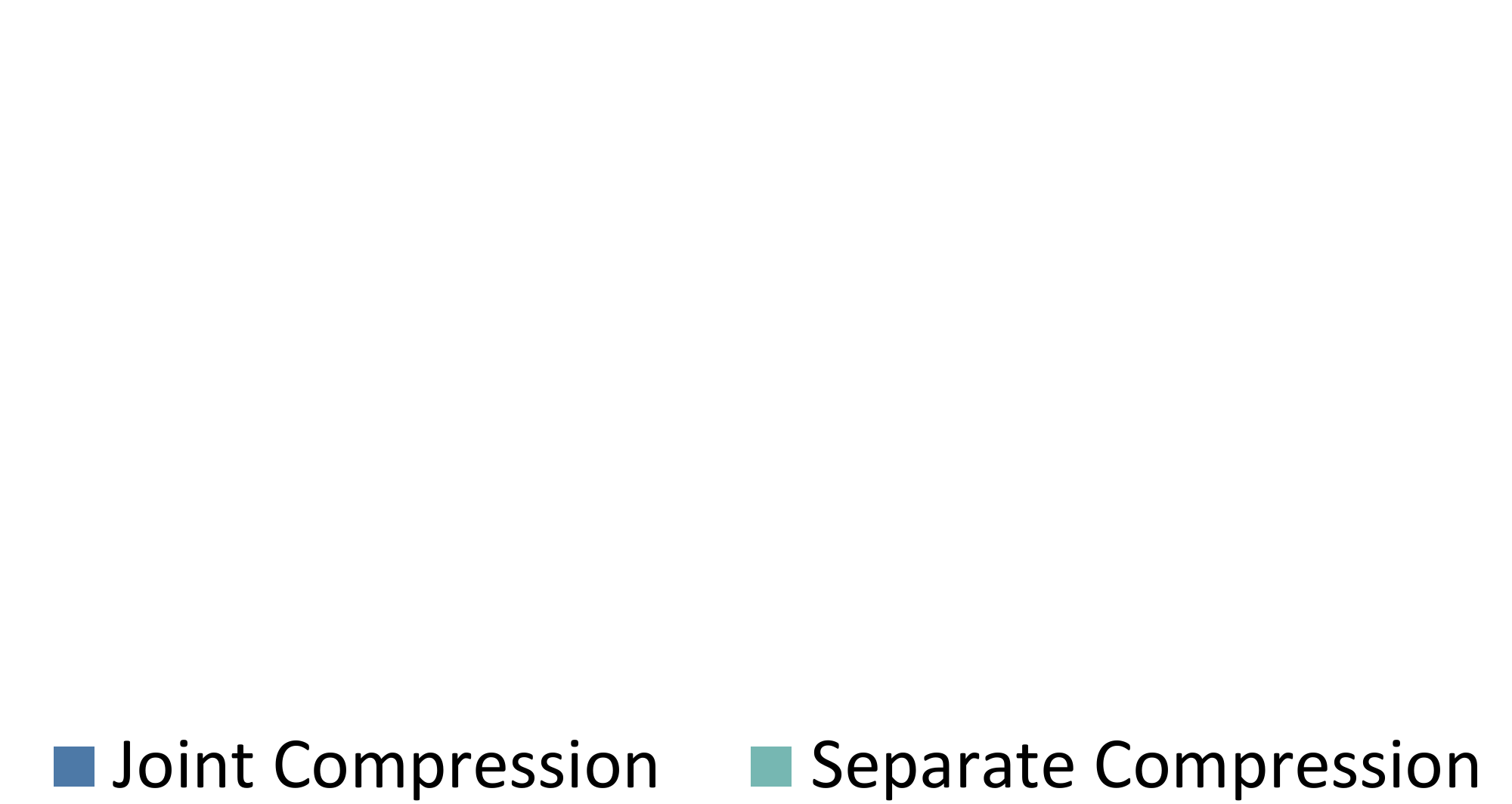}
\end{minipage}
\subfigure[\columnwidth/2][Read throughput]{
  \includegraphics[width=0.9\columnwidth/2, trim=0 1.2cm 5.25cm 0, clip=true]{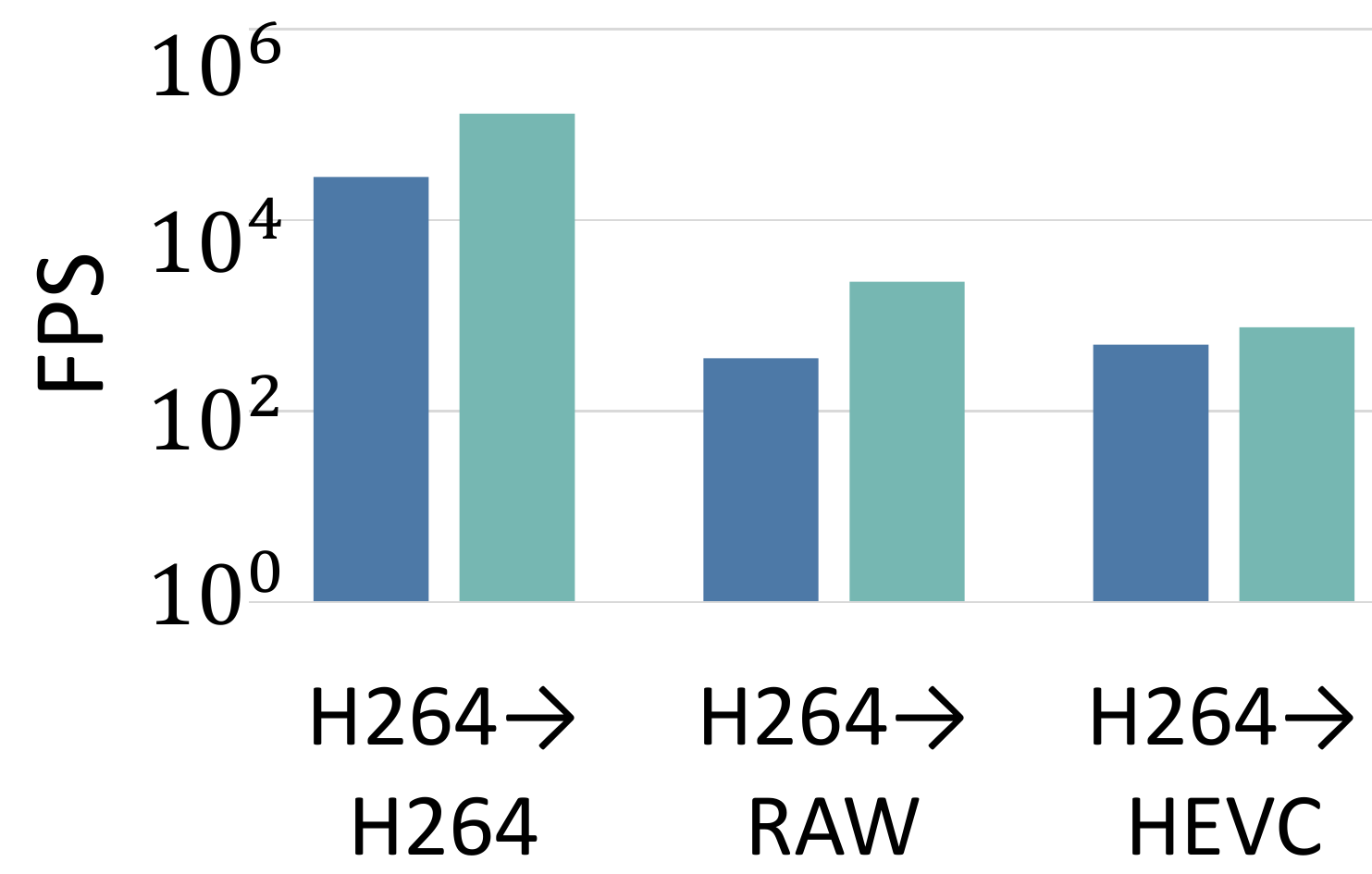}
  \label{subfig:joint-reads}
}%
\subfigure[\columnwidth/2][Write throughput]{
  \includegraphics[width=0.9\columnwidth/2, trim=0 1.2cm 5.25cm 0, clip=true]{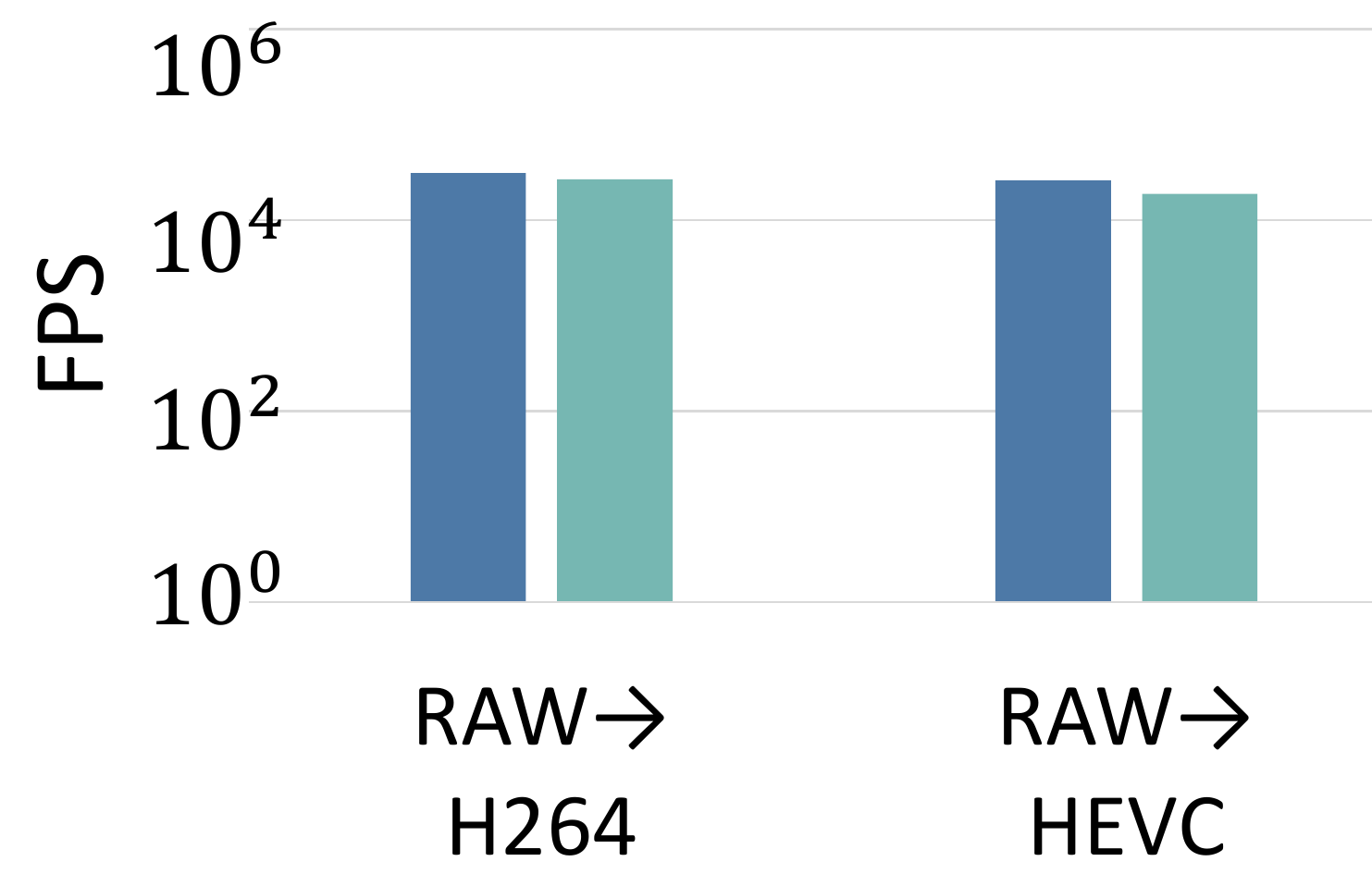}
  \label{subfig:joint-writes}
}  
\caption{Joint compression throughput.}
\label{fig:joint-throughput}
\end{figure}
}

\newcommand{\cacheEvictionAndJointSizeFigure}{
\begin{figure}[t]
\begin{minipage}{\linewidth}
\begin{minipage}{\columnwidth/2 - 0.25em}
  \centering
  \includegraphics[width=\columnwidth - 0.25em]{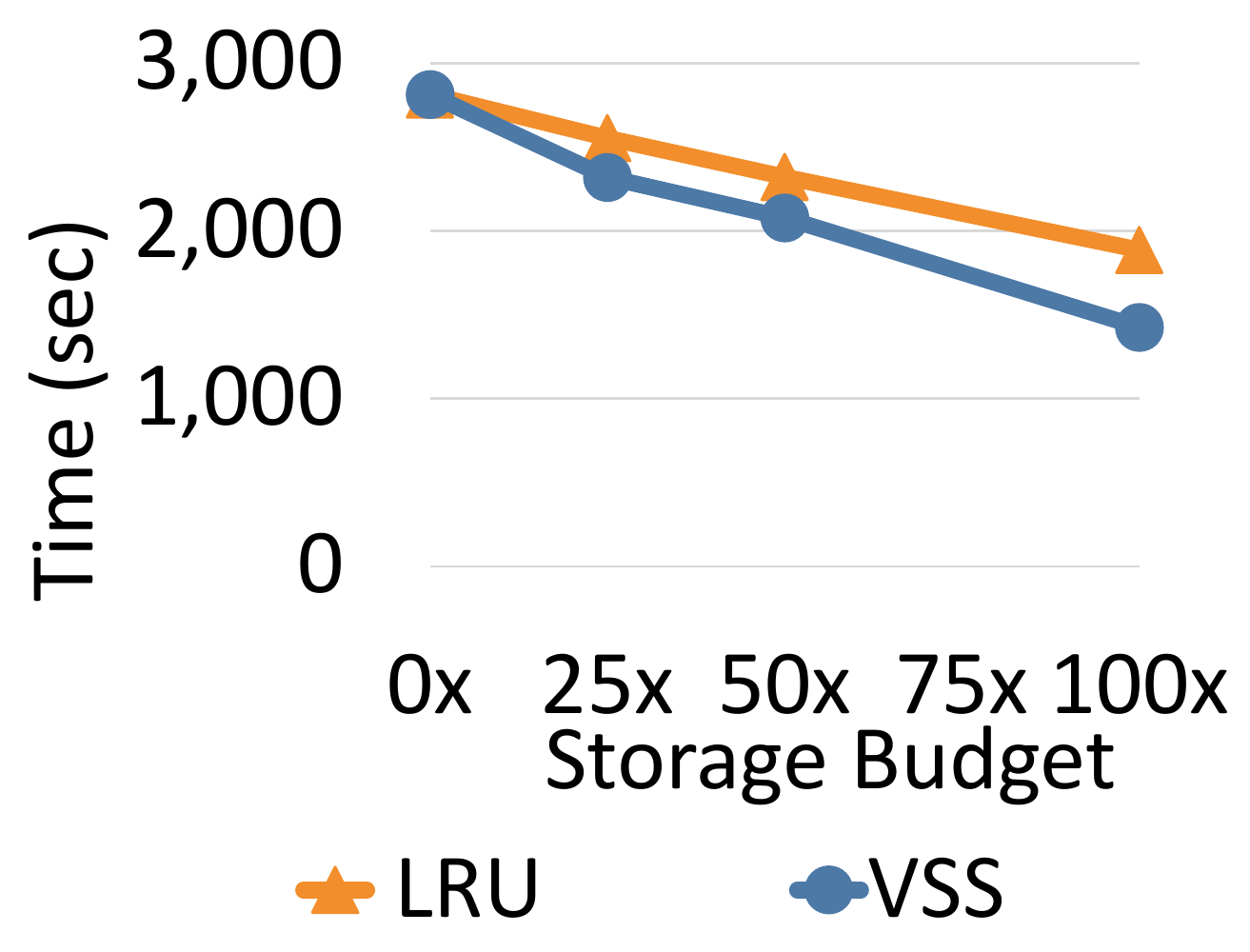}
  \caption{Read runtime by cache eviction policy.}
  \label{fig:eviction}
\end{minipage}\hspace{0.5em}%
\begin{minipage}{\columnwidth/2 - 0.25em}
  \centering
  \includegraphics[width=\columnwidth - 0.25em]{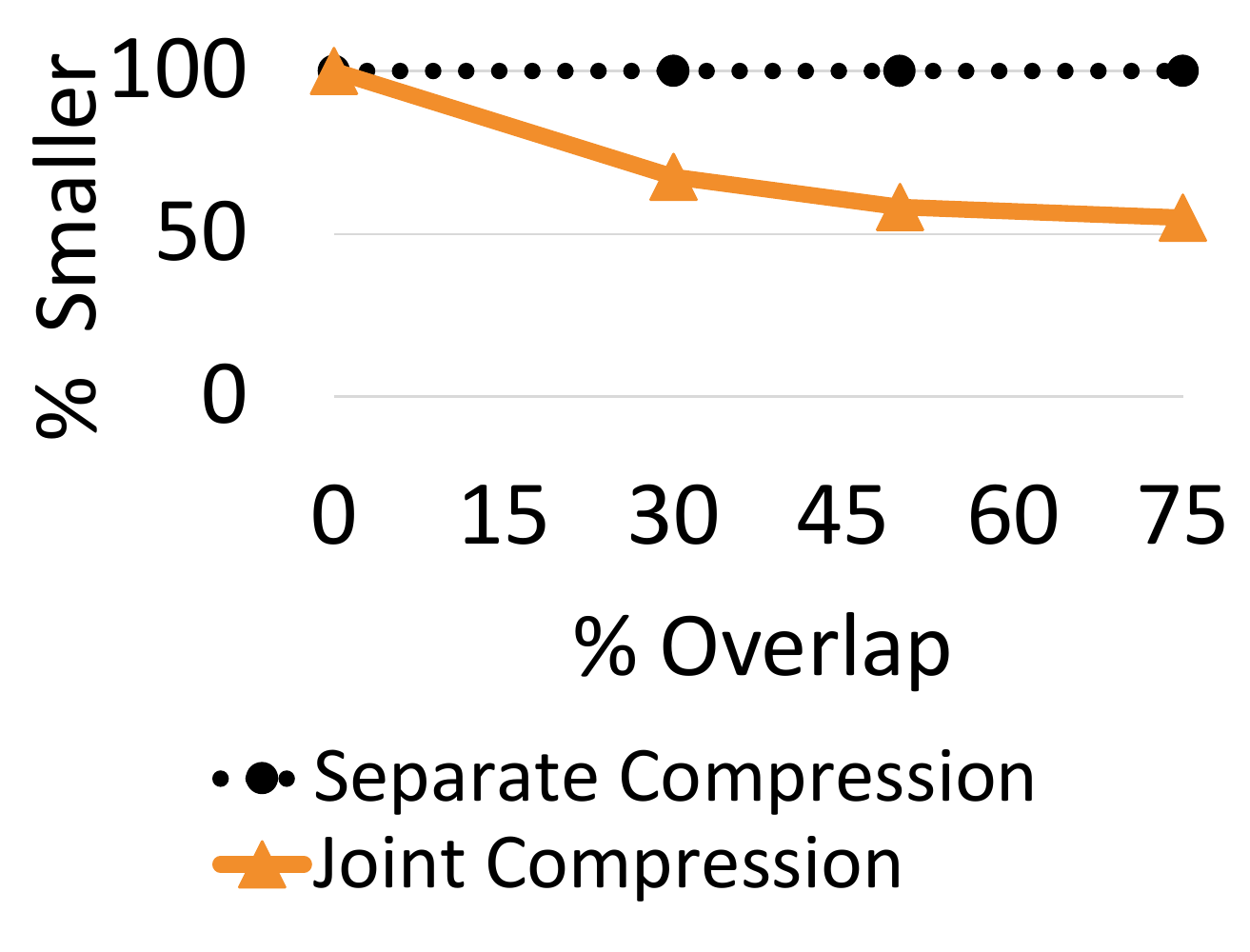}
  \caption{\squeezeevenmore{Joint compressed vs separate video.}}
  \label{fig:joint-size}
\end{minipage}
\end{minipage}
\end{figure}
}

\newcommand{\jointCompressionPerformanceFigure}{
  \includegraphics[width=\columnwidth]{figures/jointcompression.pdf}
}

\newcommand{\jointCompressionSizeFigure}{
}

\newcommand{\vssMotivationFigure}{
\begin{figure}[t]
  \centering
  \includegraphics[width=\columnwidth]{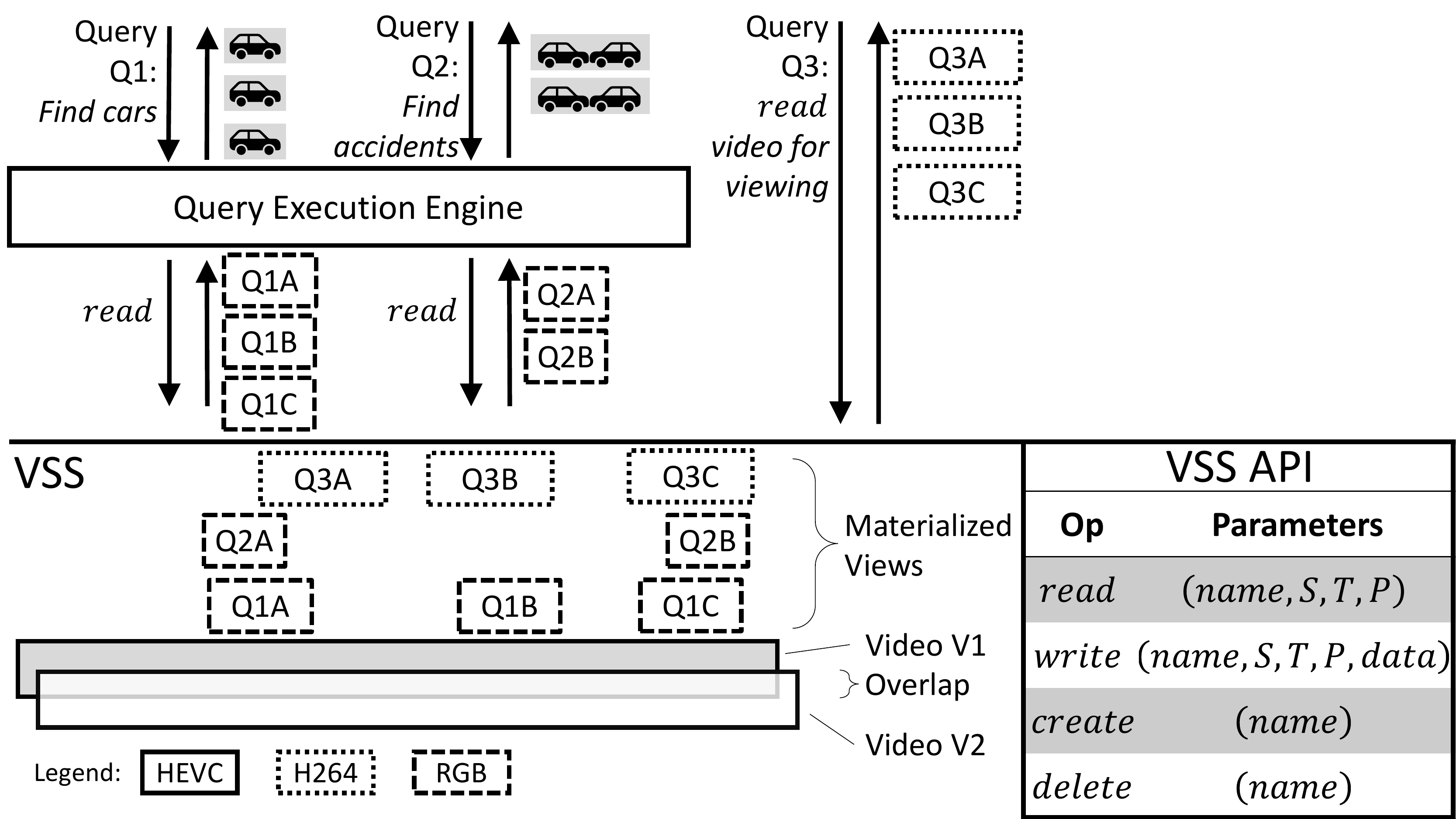}
  \caption{\squeezemore{\name overview \& API. Reads and writes require specification of spatial ($S$; resolution, region of interest), temporal ($T$; start/end time, frame rate), and physical ($P$; frame layout, compression codec, quality) parameters.}}
  \label{fig:vss-motivation}
\end{figure}
}

\newcommand{\vssTemplates}{
\begin{table}[t]
\centering
\caption{\squeezemore{\name operations.  Both reads and writes require specification of spatial ($S$; resolution, region of interest), temporal ($T$; start and end time, frame rate), and physical ($P$; frame layout, compression codec, quality) parameters.} \alvin{I suggest merging this with fig 1 to save space}}
\begin{tabularx}{\columnwidth}{lX}
  \toprule
  Op & Parameters \\
  \midrule

  $read$   & $\big(name, S, T, P\big)$ \\
  $write$  & $\big(name, S, T, P, data\big)$ \\
  $create$ & $\big(name, budget\big)$ \\
  $delete$ & $\big(name\big)$ \\
  $pin$   & $\big(name, S, T, P\big)$ \\
  $unpin$ & $\big(name, S, T, P\big)$ \\

  \bottomrule
\end{tabularx}
\label{table:vss-templates}
\end{table}
}

\newcommand{\datasetsTable}{
\begin{table}[t]
\centering
\caption{Datasets used to evaluate \name}
\setlength\tabcolsep{4pt}
\begin{tabularx}{\columnwidth}{lcrrr}
  \toprule
  \multirow{2}{*}{Dataset} & 
  \multirow{2}{*}{Resolution} & 
  \multirow{2}{*}{\# Frames} & 
  Compressed \\
  & & & Size (MB)\hspace{4pt} \\
  \midrule
  Robotcar       & $1280{\times}960$ & 7,494      & 120   \\ %
  Waymo         & $1920{\times}1280$  & 398      & 7     \\ %
  VisualRoad 1K-30\% & $960{\times}540$   & 108k & 224   \\ %
  VisualRoad 1K-50\% & $960{\times}540$   & 108k & 232   \\
  VisualRoad 1K-75\% & $960{\times}540$   & 108k & 226   \\
  VisualRoad 2K-30\% & $1920{\times}1080$ & 108k &  818  \\ %
  VisualRoad 4K-30\% & $3840{\times}2160$ & 108k & 5,500 \\ %
  \midrule
  \bottomrule
\end{tabularx}
\label{table:datasets}
\end{table}
}

\newcommand{\jointCompressionQualityTable}{
\begin{table}[t]
\centering
\footnotesize
\caption{Joint compression recovered quality}
\begin{tabularx}{\columnwidth}{p{2.25cm}ccc}
  \toprule
  \multirow{3}{*}{\small Dataset} & \multicolumn{2}{c}{\small \hspace{-0.5em}Quality (PSNR)} & \small \revision{Fragments} \\
  & \small Unprojected & \small Mean & \small \revision{Admitted (\%)} \\
  & Left\hspace{0.05em}/\hspace{0.05em}Right & Left\hspace{0.05em}/\hspace{0.05em}Right & \revision{Unprojected~/~Mean} \\
  \midrule
  \small Robotcar      & \small 350~/~24 & \small 30~/~27 & \small \revision{36~/~64}\\
  \small Waymo         & \small 352~/~29 & \small 32~/~30 & \small \revision{39~/~68}\\
  \small VRoad-1K-30\% & \small 359~/~30 & \small 31~/~30 & \small \revision{46~/~80} \\
  \small VRoad-1K-50\% & \small 358~/~28 & \small 29~/~29 & \small \revision{41~/~72} \\
  \small VRoad-1K-75\% & \small 348~/~24 & \small 30~/~28 & \small \revision{44~/~68} \\
  \small VRoad-2K-30\% & \small 352~/~30 & \small 30~/~30 & \small \revision{52~/~82} \\
  \small VRoad-4K-30\% & \small 360~/~30 & \small 29~/~30 & \small \revision{54~/~78} \\

  \bottomrule
\end{tabularx}
\label{table:joint-compression-quality}
\end{table}
}

\newcommand{\writesSizesTable}{
\begin{table}[t]
\centering
\caption{Joint compressed size by dataset (in megabytes)}
\begin{tabularx}{\columnwidth}{lrrrrr}
  \toprule
  \multirow{2}{*}{Dataset}       & \multicolumn{3}{c}{Size (MB)} \\
  & \name  & VStore & OpenCV \\
  \midrule
  Robotcar       & 26 & 123 (\phantom{0}$4.7\times$) \\
  Waymo         & \phantom{0}6  & 7 (\phantom{0}$1.2\times$) \\
  VisualRoad 1K-30\% & \phantom{0}2.7 & 244 ($90.4\times$) \\
  VisualRoad 1K-50\% \\
  VisualRoad 1K-75\% \\

     \bottomrule
\end{tabularx}
\label{table:writesizes}
\end{table}
}

\newcommand{\writesTimesTable}{
\begin{table}[t]
\centering
\caption{Write times in seconds by dataset}
\begin{tabularx}{\columnwidth}{llllll}
  \toprule
  \multirow{2}{*}{Dataset}       & \multicolumn{4}{c}{Write Time (sec)} \\
   & \name  & VStore & OpenCV & Local FS \\

  \midrule
  \hspace{-0.3em}\textbf{Raw Input} \\
  ~Robotcar       & 298 & 26 \\
  ~Waymo         & \phantom{00}8 & \phantom{0}3 \\
  ~VisualRoad 1K & \phantom{0}26 \\
  ~VisualRoad 2K & \\
  ~VisualRoad 4K & \\

  \midrule
  \hspace{-0.3em}\textbf{\avc Input} \\
  ~Robotcar       & & \phantom{0}1.9 & & 1.4 \\
  ~Waymo         & & \phantom{0}0.4 & & 0.2 \\
  ~VisualRoad 1K & & \phantom{0}4.7 \\
  ~VisualRoad 2K & & 13.8 \\
  ~VisualRoad 4K & 11.1  & 140 \\

     \bottomrule
\end{tabularx}
\label{table:writetimes}
\end{table}
}

\newcommand{\visualRoadMicrobenchmarkTable}{
\begin{tabularx}{\columnwidth}{lcX}
  \toprule
  Query & \name & ext4 \\
  \midrule

  Q1 & ? & ? \\
  \vdots \\ 
  Q6(b) & ? & ? \\
   
  \bottomrule
\end{tabularx}
}

\newcommand{\visualRoadCompositeTable}{
\begin{tabularx}{\columnwidth}{lcX}
  \toprule
  Query & \name & ext4 \\
  \midrule

  Q7 & ? & ? \\
  \vdots \\ 
  Q10 & ? & ? \\
   
  \bottomrule
\end{tabularx}
}

\newcommand{\opencvApplicationTable}{
\begin{tabularx}{\columnwidth}{lcX}
  \toprule
  Application & \name & ext4 \\
  \midrule

  Application 1 & ? & ? \\
  \vdots \\ 
  Application $n$ & ? & ? \\
   
  \bottomrule
\end{tabularx}
}

\section{Introduction}
\label{sec:introduction}

\vssMotivationFigure

\squeeze{ 
The volume of video data captured and processed is rapidly increasing: YouTube receives more than 400 hours of uploaded video per minute~\cite{youtube-stats}, and 
more than six million closed-circuit television cameras populate the United Kingdom, collectively amassing an estimated 7.5 petabytes of video per day~\cite{cloudview}.
More than 200K body-worn cameras are in service~\cite{hyland2018body}, collectively generating almost a terabyte of video per day~\cite{yates2018body}.  %
}

\revision{
To support this video data deluge, many systems and applications have emerged to ingest, transform, and reason about such data~\cite{DBLP:conf/cloud/LuCK16,DBLP:journals/pvldb/HaynesMABCC18,DBLP:conf/sigcomm/JiangABSS18,DBLP:journals/tog/PomsCHF18,DBLP:journals/pvldb/KangBZ19,DBLP:conf/osdi/HsiehABVBPGM18,DBLP:journals/pvldb/KangEABZ17,DBLP:conf/nsdi/ZhangABPBF17}.  Critically, however, most of these systems lack efficient storage managers. 
}
\squeezemore{
They focus on query execution for a video that is already decoded and loaded in memory~\cite{DBLP:journals/pvldb/KangEABZ17,DBLP:journals/pvldb/KangBZ19,DBLP:conf/osdi/HsiehABVBPGM18} or treat video compression as a black box~\cite{DBLP:conf/cloud/LuCK16,DBLP:conf/sigcomm/JiangABSS18,DBLP:conf/nsdi/ZhangABPBF17} (cf.~\cite{DBLP:journals/pvldb/HaynesMABCC18,DBLP:journals/tog/PomsCHF18}). In practice, of course, videos are stored on disk, and the cost of reading and decompressing is high relative to subsequent processing~\cite{DBLP:journals/pvldb/HaynesMABCC18,daum2020tasm}, e.g., constituting more than 50\% of total runtime~\cite{kang2020jointly}.  
The result is a performance plateau limited by Amdahl's law, where an emphasis on post-decompression performance might yield impressive results in isolation, but ignores the diminishing returns when performance is evaluated end-to-end.
}

\revision{
In this paper, we develop \name, a \textit{video storage system} designed to serve as storage manager beneath a video data management system or video processing application (collectively VDBMSs).  
Analogous to a storage and buffer manager for relational data, \name assumes responsibility for storing, retrieving, and caching video data.
It frees higher-level components to focus on application logic, while \name optimizes the low-level performance of video data storage.
As we will show, this decoupling dramatically speeds up video processing queries and decreases storage costs.
\name does this by addressing the following three challenges:
}

\squeeze{
First, modern video applications commonly issue multiple queries over the same (potentially overlapping) video regions and build on each other in different ways (e.g., \autoref{fig:vss-motivation}).
Queries can also vary video resolution and other characteristics (e.g., the SMOL system rescales video to various resolutions~\cite{kang2020jointly} and Chameleon dynamically adjusts input resolution~\cite{DBLP:conf/sigcomm/JiangABSS18}).
Such queries can be dramatically faster with an efficient storage manager that maintains and evolves a cache of video data,
each differently compressed and encoded.
}

\squeezemore{ 
Second, if the same video is queried using multiple systems such as via a VDBMS optimized for simple select and aggregate queries~\cite{DBLP:journals/pvldb/KangBZ19} and a separate vision system optimized for reasoning about complex scenes~\cite{DBLP:conf/cvpr/TsaiDMSF19} (e.g., \autoref{fig:vss-motivation}), then the video file may be requested at different resolutions and frame rates and using different encodings. Having a single storage system that encapsulates all such details and provides a unified query interface makes it seamless to create---and optimize---such federated workflows. 
While some systems have attempted to mitigate this by making multiple representations available to developers~\cite{DBLP:conf/fast/VeeraraghavanFNN10,DBLP:conf/eurosys/XuBL19}, they expensively do so for entire videos even if only small subsets (e.g., the few seconds before and after an accident) are needed in an alternate representation.
}

Third, many recent applications analyze large amounts of 
video data with overlapping fields of view and proximate locations.  For example, traffic monitoring networks often have multiple cameras oriented toward the same intersection and autonomous driving and drone applications come with multiple overlapping sensors that capture nearby video.  Reducing the redundancies that occur among these sets of physically proximate or otherwise similar video streams is neglected in all modern VDBMSs.
\squeeze{This is because of the substantial difficulties involved: systems (or users) need to consider the locations, orientations, and fields of view of each camera to identify redundant video regions; measure overlap, jitter, and temporally align each video; and ensure that deduplicated video data can be recovered with sufficient quality.
Despite these challenges, and as we show herein, deduplicating overlapping video data streams offers opportunities to greatly reduce storage costs.}

\revision{
\name addresses the above challenges. As a storage manager, it exposes a simple interface where VDBMSs read and write videos using \name's API (see \autoref{fig:vss-motivation}).  Using this API, systems write video data in any format, encoding, and resolution---either compressed or uncompressed---and \name manages the underlying compression, serialization, and physical layout on disk.  When these systems subsequently read video---once again in any configuration and by optionally specifying regions of interest and other selection criteria---\name automatically identifies and leverages the most efficient methods to retrieve and return the requested data.
}

\name deploys the following optimizations and caching mechanisms to improve read and write performance.
First, rather than storing video data on disk as opaque, monolithic files, \name decomposes video into sequences of contiguous, independently-decodable sets of frames. 
In contrast with previous systems that treat video as static and immutable data, \name applies transformations at the granularity of these sets of frames, freely transforming them as needed to satisfy a read operation.  For example, if a query requests a video region compressed using a different codec, \name might elect to cache the transcoded subregion and delete the original.

\squeeze{
As \name handles requests for video over time, it maintains a per-video on-disk collection of materialized views that is populated passively as a byproduct of read operations.
When a VDBMS performs a subsequent read, \name leverages a minimal-cost subset of these views to generate its answer. Because these materialized views 
can arbitrarily overlap and have complex interdependencies, finding the least-cost set of views is non-trivial. \name uses a satisfiability modulo theories (SMT) solver to identify the best views to satisfy a request.
\name prunes stale views by selecting those least likely to be useful in answering subsequent queries.  Among equivalently useful views, \name optimizes for video quality and defragmentation.
}

Finally, \name reduces the storage cost of redundant video data collected from physically proximate cameras.  \squeeze{It does so by deploying a \textit{joint compression} optimization that identifies overlapping regions of video and
stores these regions only once. The key challenge lies in efficiently identifying potential candidates for joint compression in a large database of videos. Our approach identifies candidates efficiently without requiring any metadata specification. To identify video overlap, \name incrementally fingerprints video fragments (i.e., it produces a feature vector that robustly characterizes video regions) and, using the resulting fingerprint index, searches for likely correspondences between pairs of videos.  It finally performs a more thorough comparison between likely pairs.}

In summary, we make the following contributions: 
\begin{itemize}[leftmargin=*,noitemsep,topsep=0em]

    \item We design a new storage manager for video data that leverages the fine-grained physical properties of videos to improve application performance (\autoref{sec:architecture}). 

    \item We develop a novel technique to perform reads by selecting from potentially many materialized views to efficiently produce an output while maintaining the quality of the resulting video data (\autoref{sec:query-answering}).

    \item We develop a method to optimize the storage required to persist videos that are highly overlapping or contain similar visual information, an indexing strategy to identify such regions (\autoref{sec:joint-compression}), and a protocol for caching multiple versions of the same video (\autoref{sec:storage-budget}).

\end{itemize}

\noindent
We evaluate \name against existing video storage techniques and show that it can reduce video read time by up to 54\% and decrease storage requirements by up to 45\% (\autoref{sec:evaluation}).

\section{\name Overview}
\label{sec:architecture}

\revision{
\squeeze{Consider an application that monitors an intersection for automobiles associated with missing children or adults with dementia.}  A typical implementation would first ingest video data from multiple locations around the intersection.  It would then index regions of interest, typically by decompressing and converting the entire video to an alternate representation suitable for input to a machine learning model trained to detect automobiles. 
Many video query processing systems provide optimizations that accelerate this process~\cite{DBLP:journals/pvldb/KangBZ19,DBLP:conf/sigmod/LuCKC18,DBLP:conf/eurosys/XuBL19}.
Subsequent operations, however, might execute more specific queries only on the regions that have automobiles.  For example, if a red vehicle is missing, a user might issue a query to identify all red vehicles in the dataset.
Afterward, a user might request and view all video sequences containing only the likely candidates.  \squeeze{This might involve further converting relevant regions to a representation compatible with the viewer (e.g., at a resolution compatible with a mobile device or compressed using a supported codec).  We show the performance of this application under \name in \autoref{sec:evaluation}.}
}

While today's video processing engines perform optimizations for operations over entire videos (e.g., the indexing phase described above), their storage layers provide little or no support for subsequent queries over the results
(even dedicated systems such as quFiles~\cite{DBLP:conf/fast/VeeraraghavanFNN10} or VStore~\cite{DBLP:conf/eurosys/XuBL19}
transcode entire videos, even when only a few frames are needed).
Meanwhile, when the above application uses \name to read a few seconds of low-resolution, uncompressed video data to find frames containing automobiles,
it can delegate responsibility to \name for efficiently producing the desired frames.  This is true even if the video is streaming or has not fully been written to disk.

Critically, \name automatically selects the most efficient way to generate the desired video data in the requested format and region of interest (ROI) based on the original video and cached representations.
\squeeze{
Further, to support real-time streaming scenarios, writes to \name are non-blocking
and
users may query prefixes of ingested video data without waiting on the entire video to be persisted.
}

\autoref{fig:vss-motivation} summarizes the set of \name-supported operations. These operations are over \textit{logical videos}, which \name executes to produce or store fine-grained \textit{physical video} data.  
Each operation involves a point- or range-based scan or insertion over a single logical video source. \name allows constraints on combinations of temporal ($T$), spatial ($S$), and physical ($P$) parameters.  Temporal parameters include start and end time interval ($[s, e]$) and frame rate ($f$); spatial parameters include resolution ($r_x \times r_y$) and region of interest ($[x_0..x_1]$ and $[y_0..y_1]$); and physical parameters $P$ include physical frame layout ($l$; e.g., \yuvfourtwozero, \yuvfourtwotwo), compression method ($c$; e.g., \hevc), and quality (to be discussed in \autoref{sec:quality-model}).

\vssPhysicalOrganization

Internally, \name arranges each written physical video as a sequence of entities called \textit{groups of pictures (GOPs)}.  
Each GOP is composed of a contiguous sequence of frames in the same format and resolution.  A GOP may include the full frame extent or be cropped to some ROI and may contain raw pixel data or be compressed.  Compressed GOPs, however, are constrained such that they are independently decodable and take no data dependencies on other GOPs.

Though a GOP may contain an unbounded number of frames, video compression codecs typically fix their size to a small, constant number of frames (30--300) and \name accepts as-is ingested compressed GOP sizes (which are typically less than 512kB).
\squeeze{
For uncompressed GOPs, our prototype implementation automatically partitions video data into blocks of size $\le 25$MB (the size of one \rgb 4K frame), or a single frame for resolutions that exceed this threshold.
}

\autoref{figure:vss-physical} illustrates the internal physical state of \name.
In this example, \name contains a single logical video \texttt{traffic} with two physical representations (one \hevc at $1920\times 1080$ resolution and 30 frames per second, and a 60-second variant at $960\times 540$ resolution).
\name has stored the GOPs associated with each representation as a series of separate files (e.g., $\texttt{\small traffic/1920x1080r30.hevc/1}$).  It has also constructed a non-clustered temporal index that maps time to the file containing associated visual information. This level of detail is invisible to applications, which access \name only through the operations summarized in \autoref{fig:vss-motivation}.

\section{Data Retrieval from \name}
\label{sec:query-answering}

\queryAnsweringFigure

\squeezemore{
As mentioned, \name internally represents a logical video as a collection of materialized physical videos. When executing a read, \name produces the result using one or more of these views.
}

\squeeze{
Consider a simplified version of the application described in \autoref{sec:architecture}, where a single camera has captured 100 minutes of 4K resolution, \hevc-encoded video, and written it to \name using the name $V$. The application first reads the entire video and applies a computer vision algorithm that identifies two regions (at minutes 30--60 and 70--95) containing automobiles.
The application then retrieves those fragments compressed using \avc to transmit to a device that only supports this format. As a result of these operations, \name now contains the original video ($m_0$) and the cached versions of the two fragments ($m_1, m_2$) as illustrated in \Cref{subfig:query-answering-views}. The figure indicates the labels $\{m_0, m_1, m_2\}$ of the three videos, their spatial configuration (\textsc{4k}), start and end times (e.g., $[0, 100]$ for $m_0$), and physical characteristics (\hevc or \avc). 
}

Later, a first responder on the scene views a one-hour portion of the recorded video on her phone, which only has hardware support for \avc decompression.  
To deliver this video, the application executes $read(V, \textsc{4k}, [20, 80], \avc)$, which, as illustrated by the arrow in \Cref{subfig:query-answering-views}, requests video $V$ at \textsc{4k} between time $[20, 80]$ compressed with $\avc$. 

\name responds by first identifying subsets of the available physical videos that can be leveraged to produce the result.  For example, \name can simply transcode $m_0$ between times $[20, 80]$.  Alternatively, it can transcode $m_0$ between time $[20, 30]$ and $[60, 70]$, $m_1$ between $[30, 60]$, and $m_2$ between $[70, 80]$.  \squeeze{
The latter plan is the most efficient since $m_1$ and $m_2$ are already in the desired output format (\avc), hence \name need not incur high transcoding costs for these regions. \autoref{subfig:query-answering-fragments} shows the different selections that \name might make to answer this read.  
Each \textit{physical video fragment} $\{f_1, .. f_6\}$ in \autoref{subfig:query-answering-fragments} represents a different region that \name might select.
Note that \name need not consider other subdivisions---for example by subdividing $f_5$ at $[30, 40]$ and $[40, 60]$---since $f_5$ being cheaper at $[30, 40]$ implies that it is at $[40, 60]$ too. 
}

To model these transcoding costs, \name employs a \textit{transcode cost model} $c_t(f, P, S)$ that 
represents the cost of converting a physical video fragment $f$ into a target spatial and physical format $S$ and $P$.  The selected fragments must be of sufficient quality, which we model using a \textit{quality model} $u(f, f')$ and reject fragments of insufficient quality.  We introduce these models in the following two subsections.

\subsection{Cost Model}
\label{sec:cost-model}

We first discuss how \name selects fragments for use in performing a read operation using its cost model.  In general, given a $read$ operation and a set of physical videos, 
\name must first select fragments that cover the desired spatial and temporal ranges.  To ensure that a solution exists, \revision{\name maintains a \textit{cover} of the initially-written video $m_0$ consisting of  physical video fragments with quality equal to the original video (i.e., $u(m_0, f) \ge \tau$).  Our prototype sets a threshold $\tau=40$dB, which is considered to be lossless. See \autoref{sec:quality-model} for details.
}
\name also returns an error for reads extending outside of the temporal interval of $m_0$.

Second, when the selected physical videos temporally overlap, \name must resolve \textit{which} physical video fragments to use in producing the answer
in a way that minimizes the total conversion cost of the selected set of video fragments.
This problem is similar to materialized view selection~\cite{DBLP:journals/vldb/Halevy01}.  \squeezeevenmore{Fortunately, a \name read is far simpler than a general database query, and in particular is constrained to a small number of parameters with point- or range-based predicates.}

We motivate our solution by continuing our example from \Cref{subfig:query-answering-views}.  
First, observe that the collective start and end points of the physical videos form a set of \textit{transition points} where \name can switch to an alternate physical video.
 In \Cref{subfig:query-answering-views}, the transition times include those in the set $\{30, 60, 70\}$, and we illustrate them in \Cref{subfig:query-answering-fragments} by partitioning the set of cached materialized views at each transition point.  \shepherd{\name ignores fragments that are outside the read's temporal range, since they do not provide information relevant to the read operation.}

Between each consecutive pair of transition points, \name must choose exactly one physical video fragment.  In \Cref{subfig:query-answering-fragments}, we highlight one such set of choices that covers the read interval.
Each choice of a fragment comes with a cost (e.g., $f_1$ has cost $32$), derived using a cost formula given by $c_t(f, P, S) = \alpha(f_S, f_P, S, P) \cdot \vert f \vert$.  This cost is proportional to the total number of pixels $\vert f \vert$ in fragment $f$
scaled by $\alpha(S, P, S', P')$, which is the normalized cost of transcoding a single pixel from spatial and physical format $(S, P)$ into format $(S', P')$.  \squeezemore{For example, using fragment $m_1$ in \autoref{fig:query-answering} requires transcoding from physical format $P=\hevc$ to $P'=\avc$ with no change in spatiotemporal format (i.e., $S=S'$).
}

\revision{
During installation, \name computes the domain of $\alpha$ by executing the \texttt{vbench} benchmark~\cite{DBLP:conf/asplos/LottariniRCKRSW18} on the installation hardware, which produces per-pixel transcode costs for a variety of resolutions and codecs.  
For resolutions not evaluated by \texttt{vbench}, \name approximates $\alpha$ by piecewise linear interpolation of the benchmarked resolutions.  
}

\lookbackCostFigure

\shepherd{\name must also consider the data dependencies between frames.} \revision{Consider the illustration in \autoref{figure:lookback-cost}, which shows the frames within a physical video with their data dependencies indicated by directed edges.  If \name wishes to use a fragment at the frame highlighted in red, it must first decode all of the red frame's {\it dependent frames}, denoted by the set $\Delta$ in \autoref{figure:lookback-cost}.  This implies that the cost of transcoding a frame depends on \textit{where} within the video it occurs, and whether its dependent frames are \textit{also} transcoded.
}

To model this, we introduce a \textit{look-back cost} $c_l(\Omega, f)$ 
that gives the cost of decoding the set of frames $\Delta$ on which fragment $f$ depends 
\textit{if they have not already been decoded}, meaning that they are not in the set of previously selected frames $\Omega$.  As illustrated in \autoref{figure:lookback-cost}, these dependencies come in two forms: independent frames $\mathrm{A} \subseteq \Delta$ (i.e., frames with out-degree zero in our graphical representation) which are larger in size but less expensive to decode, and the remaining dependent frames $\Delta - \mathrm{A}$ (those with outgoing edges) which are highly compressed but have more expensive decoding dependencies between frames.  
We approximate these per-frame costs using estimates from Costa et al.~\cite{DBLP:conf/iccel/CostaAC18}, which empirically concludes that dependent frames are approximately $45\%$ more expensive than their independent counterparts.
We therefore fix $\eta=1.45$ and formalize look-back 
cost as $c_l(\Omega, f) = \vert \mathrm{A} - \Omega \vert + \eta \cdot \vert (\Delta - \mathrm{A}) - \Omega \vert$.

\squeezemore{
To conclude our example, observe that our goal is to choose a set of 
physical video fragments that cover the queried spatiotemporal range, do not temporally overlap, and minimize the decode and look-back cost of selected fragments.  In \Cref{subfig:query-answering-fragments}, of all the possible paths, the one highlighted in gray minimizes this cost. These characteristics collectively meet the requirements identified at the beginning of this section.
}

Generating a minimum-cost solution using this formulation
requires jointly optimizing both look-back cost $c_l$ and transcode cost $c_t$,
where each fragment choice affects the dependencies (and hence costs) of future choices.  These dependencies make the problem not solvable in polynomial time, and \name employs an SMT solver~\cite{DBLP:conf/tacas/MouraB08} to generate an optimal solution.  Our embedding constrains frames in overlapping fragments so that only one is chosen, selects combinations of regions of interest (ROI) that spatially combine to cover the queried ROI, and uses information about the locations of independent and dependent frames in each physical video to compute the cumulative decoding cost due to both transcode and look-back for any set of selected fragments.  
We compare this algorithm to a dependency-na\"ive greedy baseline in \autoref{sec:evaluation:read}.

\subsection{Quality Model}
\label{sec:quality-model}

Besides efficiency, \name must also ensure that the quality of a result has sufficient fidelity.  For example, using a heavily \textit{downsampled} (e.g., $32\times32$ pixels) or \textit{compressed} (e.g., at a 1Kbps bitrate) physical video to answer a read requesting \textsc{4k} video is likely to be unsatisfactory.
\name tracks quality loss from both sources using a quality model $u(f_0, f)$ that gives the expected quality loss of using a fragment $f$ in a read operation relative to using the originally-written video $f_0$.  
When considering using a fragment $f$ in a read, \name will reject it if the expected quality loss is below a user-specified cutoff: $u(f_0, f) < \epsilon$.  The user optionally specifies this cutoff in the read's physical parameters (see \autoref{fig:vss-motivation}); otherwise, a default threshold is used ($\epsilon=40$dB in our prototype).

The range of $u$ is a non-negative peak signal-to-noise ratio (PSNR), a common measure of quality variation based on mean-squared error~\cite{DBLP:conf/icpr/HoreZ10}.  Values ${\ge} 40$dB are considered to be lossless qualities, and ${\ge} 30$dB near-lossless. PSNR is itself defined in terms of the mean-squared error (MSE) of the $n$ pixels in frame $f_{i}$ relative to the corresponding pixels in reference frame $f_{0}$, normalized by the maximum possible pixel value $I$ (generally 255): 
\vspace{-0.5em}
\begin{equation*}
PSNR(f_i, f_0) = \frac{1}{n} \sum_{j=0}^n 10\cdot \log_{10}\big(\frac{I^2}{MSE(f_{i}[j], f_{0}[j])}\big)
\end{equation*}

\vspace{-0.5em}
As described previously, error in a fragment accumulates through two mechanisms---resampling and compression---and \name uses the sum of both sources when computing $u$.  
\revision{We next examine how \name computes error from each source.}

\revision{\textbf{Resampling error}.}  First, for downsampled error produced through a resolution or frame rate change applied to $f_0$, computing $MSE(f, f_0)$ is straightforward. 
However, \name may transitively apply these transformations to a sequence of fragments.  For example, $f_0$ might be downsampled to create $f_1$, and $f_1$ later used to produce $f_2$.  In this case, when computing $MSE(f_0, f_2)$, \name no longer has access to the uncompressed representation of $f_0$.   Rather than expensively re-decompressing $f_0$, \name instead bounds $MSE(f_0, f_{n})$ in terms of $MSE(f_0, f_1), ..., MSE(f_{n-1}, f_{n})$, which are a single real-valued aggregates stored as metadata.  
This bound is derived as follows for fragments of resolution $m\times n$:
\begingroup
\allowdisplaybreaks
\vspace{-0.5em}
\begin{align*}
{MSE}(f_0, f_2) &= \frac{1}{mn}\sum_{i=0}^{m-1}\sum_{j=0}^{n-1}(f_0^{ij} - f_2^{ij})^2 \\
&= \frac{1}{mn}\sum_{i=0}^{m-1}\sum_{j=0}^{n-1}
    \big[(f_0^{ij} - f_2^{ij})^2 + 2(f_1^{ij})^2 - 2(f_1^{ij})^2 \\
    &\phantom{====}
         + 2 f_1^{ij} f_2^{ij} - 2 f_1^{ij} f_2^{ij} 
         + 2 f_0^{ij} f_1^{ij} - 2 f_0^{ij} f_1^{ij}
         \big] \\
&= \frac{1}{mn}\sum_{i=0}^{m-1}\sum_{j=0}^{n-1}
  \begin{multlined}[t]
    \big[(f_0^{ij} - f_1^{ij})^2 + (f_1^{ij} - f_2^{ij})^2 \\
      + 2f_1^{ij}(f_2^{ij} - f_1^{ij}) - 2f_0^{ij}(f_2^{ij} - f_1^{ij})\big] \\
  \vspace{-1em}
  \end{multlined} \\
&= \begin{multlined}[t]
  MSE(f_0, f_1) + MSE(f_1, f_2) \\ 
  + \frac{2}{mn}\sum_{i=0}^{m-1}\sum_{j=0}^{n-1} \big[(f_1^{ij} - f_0^{ij}) \cdot (f_2^{ij} - f_1^{ij})\big] \\
  \vspace{-1em}
  \end{multlined} \\
&= \begin{multlined}[t]
  MSE(f_0, f_1) + MSE(f_1, f_2) \\ 
  + \frac{2}{mn} \bigg[
      \sum_{i=0}^{m-1}\sum_{j=0}^{n-1} (f_1^{ij} - f_0^{ij}) \cdot (f_2^{ij} - f_1^{ij}) \cdot I(\cdot > 0) + \\
      \sum_{i=0}^{m-1}\sum_{j=0}^{n-1} (f_1^{ij} - f_0^{ij}) \cdot (f_2^{ij} - f_1^{ij}) \cdot I(\cdot < 0)
    \bigg] \\
  \vspace{-1em}
  \end{multlined} \\
&\le \begin{multlined}[t]
  MSE(f_0, f_1) + MSE(f_1, f_2) \\ 
  + \frac{2}{mn} 
      \sum_{i=0}^{m-1}\sum_{j=0}^{n-1} (f_1^{ij} - f_0^{ij}) \cdot (f_2^{ij} - f_1^{ij}) \cdot I(\cdot > 0)
       \\
  \vspace{-1em}
  \end{multlined} \\
&\le \begin{multlined}[t]
  MSE(f_0, f_1) + MSE(f_1, f_2) \\ 
  + \frac{2}{mn} 
      \sum_{i=0}^{m-1}\sum_{j=0}^{n-1} \bigg[\frac{(f_1^{ij} - f_0^{ij}) + (f_2^{ij} - f_1^{ij})}{2}\bigg]^2 %
    \\
  \vspace{-1em}
  \end{multlined} \\
&= \begin{multlined}[t]
  MSE(f_0, f_1) + MSE(f_1, f_2) \\ 
  + \frac{1}{2mn} \bigg[
      \sum_{i=0}^{m-1}\sum_{j=0}^{n-1} (f_2^{ij} - f_0^{ij})^2 \cdot I(\cdot > 0)
    \bigg] \\
  \vspace{-1em}
  \end{multlined} \\
&\le \begin{multlined}[t]
  MSE(f_0, f_1) + MSE(f_1, f_2) + \frac{1}{2} MSE(f_0, f_2) \\
  \vspace{-1em}
  \end{multlined} \\
&= 2(MSE(f_0, f_1) + MSE(f_1, f_2)) \\
\end{align*}
\endgroup

\vspace{-2em}

Using the above formulation, \name efficiently estimates MSE for two transformations without requiring that the first fragment be available.  
Extension to transitive sequences is straightforward.

\revision{
\textbf{Compression error}.
Unlike resampling error, tracking quality loss due to lossy compression error is challenging because it cannot be calculated without decompressing---an expensive operation---and comparing the recovered version to the original input.  Instead, \name estimates compression error in terms of mean bits per pixel per second (MBPP/S), which is a metric reported during (re)compression.  
\name then estimates quality by mapping MBPP/S to the
PSNR reported by the \texttt{vbench} benchmark~\cite{DBLP:conf/asplos/LottariniRCKRSW18}, a benchmark for evaluating video transcode performance in the cloud.  To improve on this estimate, \name periodically samples regions of compressed video, computes exact PSNR, and updates its estimate.
}

\section{Data Caching in \name}
\label{sec:storage-budget}

We now describe how \name decides \textit{which} physical videos to maintain, and which to evict under low disk space conditions.  
This involves making two interrelated decisions: 
\begin{itemize}[noitemsep,topsep=0.5em]
  \setlength\itemsep{3pt}
	\setlength{\parskip}{0pt}
  \setlength{\parsep}{0pt}
	\item \squeeze{When executing a read, should \name admit the result as a new physical video for use in answering future reads?}
	\item When disk space grows scarce, which existing physical video(s) should \name discard?
\end{itemize}
To aid both decisions, \name maintains a video-specific \textit{storage budget} that limits the total size of the physical videos associated with each logical video.  The storage budget is set when a video is created in \name (see \autoref{fig:vss-motivation})
and may be specified as a multiple of the size of the initially written physical video or a fixed ceiling in bytes.  This value is initially set to an administrator-specified default ($10{\times}$ the size of the initially-written physical video in our prototype).
\revision{
\squeeze{
As described below, \name ensures a sufficiently-high quality version of the original video can always be reproduced.  It does so by maintaining a cover of fragments with sufficiently high quality (PSNR $\ge 40$dB in our prototype, which is considered to be lossless) relative to the originally ingested video.
}
}

\vssCaching
As a running example, consider the sequence of reads illustrated in \autoref{figure:caching}, which mirrors the alert application 
described in \autoref{sec:architecture}.  In this example, an application reads a low-resolution uncompressed video from \name for use with an automobile detection algorithm. \name caches the result as a sequence of three-frame GOPs (approximately 518kB per GOP).  One detection was marginal, and so the application reads higher-quality 2K video to apply a more accurate detection model.  \name caches this result as a sequence of single-frame GOPs, since each 2K \rgb frame is 6MB in size.  Finally, the application extracts two \avc-encoded regions for offline viewing.  \name caches $m_3$, but when executing the last read it determines that it has exceeded its storage budget and must now decide whether to cache $m_4$.

The key idea behind \name's cache is to logically break physical videos into ``pages.''  That is, rather than treating each physical video as a monolithic cache entry, \name targets the individual GOPs \textit{within} each physical video.  Using GOPs as cache pages greatly homogenizes the sizes of the entries that \name must consider.  
\name's ability to evict GOP pages \textit{within} a physical video differs from other variable-sized caching efforts such as those used by content delivery networks (CDNs), which make decisions on large, indivisible, and opaque entries (a far more challenging problem space with limited solutions~\cite{DBLP:journals/pomacs/BergerBH18}).

However, there are several key differences between GOPs and pages. In particular, GOPs are related to each other; i.e., (i) one GOP might be a higher-quality version of another, and (ii) consecutive GOPs form a contiguous video fragment.
These correlations make typical eviction policies like least-recently used (LRU) inefficient.  In particular, na\"{i}ve LRU might evict every other GOP in a physical video, decomposing it into many small fragments and increasing the cost of reads (which have exponential complexity in the number of fragments).  

Additionally, given multiple, redundant GOPs that are all variations of one another, ordinary LRU would treat eviction of a redundant GOP the same as any other GOP.  However, our intuition is that it is desirable to treat redundant GOPs different than singleton GOPs without such redundancy.

\revision{
Given this intuition, \name employs a modified LRU policy ($LRU_{\name}$)
that associates each fragment with a nonnegative sequence number computed using ordinary LRU offset by:
\begin{itemize}[leftmargin=1.25em,noitemsep,topsep=0em] %
  \setlength\itemsep{3pt}
  \setlength{\parskip}{0pt}
  \setlength{\parsep}{0pt}

  \item\textbf{Position ($p$).}
   \shepherd{To reduce fragmentation, \name increases the sequence number of fragments near the middle of a physical video, relative to the beginning or end.}
  For a video with $n$ fragments arranged in ascending temporal order, \name increases the sequence number of fragment $f_i$ by $p(f_i) = \min(i, n-i)$.

  \item \textbf{Redundancy ($r$).}
  \shepherd{\name decreases the sequence number of fragments that have  
  redundant or higher-quality variants.}
  To do so, using the quality cost model $u$, \name generates a $u$-ordering of each fragment $f_i$ and all other fragments that are a spatiotemporal cover of $f_i$.  
  \name decreases the sequence number of $f_i$ by its rank $r(f_i)\colon \mathbb{Z}^{0+}$ in this order
  (i.e., $r(f_i)=0$ for a fragment with no higher-quality alternatives, while $r(f_i)=n$ for a fragment with $n$ higher-quality variants).

  \item 
  \textbf{Baseline quality ($b$).}
  \shepherd{\name never evicts a fragment if it is the only fragment with quality equal to the quality of the corresponding fragment $m_0$ in the originally-written physical video. 
  To ensure this, given a set of fragments $F$ in a video, 
  \name increases the sequence number of each fragment by (our prototype sets $\tau=40$):}
  \begin{equation*}
    b(f_i) =
    \begin{cases*}
                    +\infty & if $\nexists f_j \in F\setminus f_i \ldotp u(m_0, f_j) \ge \tau$  \\
                    0 & otherwise
    \end{cases*}
  \end{equation*}
\end{itemize}

  Using the offsets described above, \name computes the sequence number of each candidate fragment $f_i$ as
   ${LRU}_{\textsc{vss}}(f_i) = LRU(f_i) + \gamma \cdot p(f_i) - \zeta \cdot r(f_i) + b(f_i)$.  
   Here weights $\gamma$ and $\zeta$ balance between position and redundancy, and our prototype weights the former ($\gamma = 2$) more heavily than the latter ($\zeta = 1$). %
   It would be a straightforward extension to expose these as parameters tunable for specific workloads.

In \autoref{figure:caching}, we show application of ${LRU}_{\textsc{vss}}$ where \name choses to evict the three-frame GOP at the beginning of $m_1$ and to cache $m_4$.  If our prototype had instead weighed $\zeta \gg \gamma$, \name would elect to evict $m_3$ since it was not recently used and is the variant with lowest quality.
}

\section{Data Compression in \name}
\label{sec:joint-compression}

As described in \autoref{sec:architecture}, when an application writes data to \name, \name partitions the written video into blocks by GOP (for compressed video data) or contiguous frames (for uncompressed video data).  \name follows the same process when caching the result of a read operation for future use.

\name employs two compression-oriented optimizations and one optimization that reduces the number of physical video fragments.  Specifically, \name (i) jointly compresses redundant data across multiple physical videos (\autoref{sec:compression-joint}); (ii) lazily compresses blocks of uncompressed, infrequently-accessed GOPs (\autoref{sec:compression-lazy}); and (iii) improves the read performance by compacting temporally-adjacent video (\autoref{sec:compaction}).

\subsection{Joint Physical Video Compression}
\label{sec:compression-joint}

\revision{
Increasingly large amounts of video content is produced from cameras that are spatially proximate with similar orientations.  For example, a bank of traffic cameras mounted on a pole will each capture video of the same intersection from similar angles.  Although the amount of ``overlapping video'' being produced is difficult to quantify, it broadly includes traffic cameras (7.5PB per day in the United Kingdom~\cite{cloudview}), body-worn cameras (>1TB per day~\cite{yates2018body}),  autonomous vehicles (>15TB per vehicle per hour~\cite{heinrich2017flash}), along with videos of  tourist locations, concerts, and political events.  
Despite the redundant information that mutually exists in these video streams, most applications treat these video streams as distinct and persist them separately to disk.
}

\squeeze{
\name optimizes the storage of these videos by reducing the redundancy between pairs of highly-similar video streams.  
\shepherd{It applies this \textit{joint compression} optimization to pairs of GOPs in different logical videos.}
\name first finds candidate GOPs to jointly compress.
Then, given a pair of overlapping GOP candidates, \name recompresses them frame-by-frame (we describe this process in \autoref{subsec:joint-compression-frame}).
For static cameras, once \name compresses the first frame in a GOP, it can reuse the information it has computed to easily compress subsequent frames in the same GOP. We describe joint compression for dynamic cameras in \autoref{subsec:joint-compression-dynamic}.  We finally describe the search process for overlapping GOPs in \autoref{subsec:joint-compression-selection}.
}

\projectionFigure

\projectionResultsFigure

\subsubsection{Joint frame compression}
\label{subsec:joint-compression-frame}
\squeeze{
\autoref{fig:projection} illustrates the joint compression process for two frames taken from a synthetic dataset (Visual Road-1K-50\%, described in \autoref{sec:evaluation}).  \Cref{subfig:projection-frame-1,subfig:projection-frame-2} respectively show the two frames with the overlap highlighted.  \autoref{subfig:projection-overlap} shows the combined regions.  
}

Because these frames were captured at different orientations, combining them is non-trivial and requires more than an isomorphic translation or rotation (e.g., the angle of the horizontal sidewalk is not aligned).  
Instead, \name estimates a homography between the two frames and a projection is used to transform between the two spaces.  As shown in \autoref{subfig:projection-overlap}, \name transforms the right frame, causing its right side to bulge vertically.  However, after it is overlaid onto the left frame, the two align near-perfectly.

\jointCompressionAlgorithm

\squeeze{
As formalized in \autoref{alg:joint-compression},
joint projection proceeds as follows.  
First, \name estimates a homography between two frames in the GOPs being compressed.  Next, it applies a feature detection algorithm~\cite{DBLP:conf/iccv/Lowe99} that identifies features that co-occur in both frames.  Using these features, it estimates the homography matrix used to transform between frame spaces.   
}

With a homography estimated, \name projects the right frame into the space of the left frame.  This results in three distinct regions: (i) a non-overlapping ``left'' region of the left frame, (ii) an overlapping region, and (iii) a ``right'' region of the right frame that does not overlap with the left.  \name splits these into three distinct regions and uses an ordinary video codec to encode each region separately and write it to disk.  %

\revision{
When constructing the overlapping region, \name applies a \textit{merge function} that transforms overlapping pixels from each overlapping region and outputs a merged, overlapping frame.  An \textit{unprojected merge} favors the unprojected frame (i.e., the left frame in \autoref{subfig:projection-overlap}), while a \textit{mean merge} averages the pixels from both input frames. 
During reads, \name reverses this process to produce the original frames.  
\autoref{fig:recovered-leftjoin} shows two such recovered frames produced using the frames shown in \autoref{fig:projection}.

Some frames stored in \name may be exact duplicates, however, for which the projection process described above introduces unnecessary computational overhead.  \name detects this case by checking whether the homography matrix would make a near-identity transform (specifically by checking $\left\vert\vert H - \mathcal{I} \vert\right\vert \le \epsilon$, where $\epsilon = \frac{1}{10}$ in our prototype).  When this condition is met, \name instead replaces the redundant GOP with a pointer to its near-identical counterpart.  %
}

\subsubsection{Dynamic \& mixed resolution cameras}
\label{subsec:joint-compression-dynamic}
For stationary and static cameras, the originally-computed homography is sufficient to jointly compress all frames in a GOP.  For dynamic cameras, however, the homography quickly becomes outdated and, in the worst case, the cameras may no longer overlap.  To guard against this, for each jointly compressed frame, \name inverts the projection process and recovers the original frame.  It then compares the recovered variant against the original using its quality model (see \autoref{sec:quality-model}).  If quality is too low ($<$24dB in our prototype), \name re-estimates the homography and reattempts joint compression, and aborts if the reattempt is also of low quality.

\revision{
For both static and dynamic cameras, \name may occasionally poorly estimate the homography between two otherwise-compatible frames.  The recovery process described above also identifies these cases.  When detected (and if re-estimation is unsuccessful), \name aborts joint compression for that pair of GOPs.  
An example of two frames where \name produced an incorrect homography is illustrated in \autoref{fig:aborted-leftjoin}.
}

\name may also identify joint compression candidates that are at dissimilar resolutions.  To handle this case, \name first upscales the lower resolution fragment to that of the higher.  It then applies joint compression as usual.

\vssCompression

\subsubsection{Selecting GOPs for joint compression}
\label{subsec:joint-compression-selection}
Thus far we have discussed how \name applies joint compression to a pair of GOPs, but not how the pairs are selected.  Since the brute force approach of evaluating all $O(n^2)$ pairs is prohibitively expensive, 
\name instead uses the multi-step process illustrated in \autoref{figure:compression-overview}.  
First, to reduce the search space, \name clusters all video fragments using their color histograms. Videos with highly distinct color histograms are unlikely to benefit from joint compression. The \name prototype implementation uses the BIRCH clustering algorithm~\cite{DBLP:conf/sigmod/ZhangRL96}, which is memory efficient, scales to many data points, and allows \name to incrementally update its clusters as new GOPs arrive.

\squeeze{
Once \name has clustered the ingested GOPs, it selects the cluster with the smallest radius and considers its constituents for joint compression.  
To do so, \name applies a modified form of the homography computation described above.  
It begins by applying the feature detection algorithm~\cite{DBLP:conf/iccv/Lowe99} from~\autoref{subsec:joint-compression-frame}.  Each feature is a spatial histogram characterizing an ``interesting region'' in the frame (i.e., a keypoint).
}  \name next looks for other GOPs in the cluster that share a large number of interesting regions.  Thus, for each GOP, \name iteratively searches for similar features (i.e., within distance $d$) located in other GOPs within the cluster. 
A correspondence, however, may be ambiguous (e.g., if a feature in GOP 1 matches to multiple, nearby features in GOP 2).  \name rejects such matches.

\squeezemore{
When \name finds $m$ or more nearby, unambiguous correspondences, it considers the pair of GOPs to be sufficiently related.  It then applies joint compression to the GOP pair as described above.  Note that the algorithm described in \autoref{subsec:joint-compression-frame} will abort if joint compressing the GOPs does not produce a sufficiently high-quality result. Our prototype sets $m=20$, requires features to be within $d=400$ (using a Euclidean metric), and disambiguates using Lowe's ratio~\cite{DBLP:journals/ijcv/Lowe04}.
}

\subsection{Deferred Compression}
\label{sec:compression-lazy}

\squeeze{
Most video-oriented applications operate over decoded video data (e.g., \rgb) that is vastly larger than its compressed counterpart (e.g., the VisualRoad-4K-30\% dataset we describe in \autoref{sec:evaluation} is 5.2TB uncompressed as 8-bit \rgb).  Caching this uncompressed video quickly exhausts the storage budget.
}

\squeeze{
To mitigate this,
\name adopts the following approach.  When a video's cache size exceeds a threshold (25\% in our prototype), \name activates its \textit{deferred compression} mode.  Thereafter when an uncompressed read occurs,
\name 
orders the video's uncompressed cache entries by eviction order.  It then losslessly compresses the \textit{last} entry (i.e., the one least likely to be evicted).  It then executes the read as usual.  
}

\squeezemore{
Our prototype uses Zstandard for lossless compression, which emphasizes speed over compression ratio (relative to more expensive codecs such as PNG or HEVC)~\cite{zstd}.
}

\squeeze{
\name performs two additional optimizations.
First, Zstandard comes with a ``compression level'' setting in the range $[1..19]$, with the lowest setting having the fastest speed but the lowest compression ratio (and vice versa).
\name linearly scales this compression level with the remaining storage budget, trading off decreased size for increased throughput.
Second, \name also compresses cache entries in a background thread 
when no other requests are being executed.
}

\subsection{Physical Video Compaction}
\label{sec:compaction}

\squeezemore{
While caching,
\name persists pairs of cached videos 
with contiguous time and the same spatial and physical configurations.
(e.g., entries at time $[0, 90]$ and $[90, 120]$).  Deferred compression may also create contiguous entries. %
}

\squeezemore{
To reduce the number of videos that need to be considered during a read, \name periodically and non-quiescently compacts pairs of contiguous cached videos and substitutes a unified representation.  It does so by 
periodically examining pairs of cached videos and, for each contiguous pair, creating hard links from the second into the first.  It then removes the second copy.
}

\section{Evaluation}
\label{sec:evaluation}

\squeeze{
We have implemented a prototype of \name in Python and \CC using CUDA~\cite{CUDA}, NVENCODE~\cite{nvenc}, OpenCV~\cite{opencv}, FFmpeg~\cite{ffmpeg}, and SQLite~\cite{sqlite}.
Our prototype adopts a no-overwrite policy and disallows updates.  We plan on supporting both features in the future.
Finally, \name does not guarantee writes are visible until the file being written is closed.
}

\textbf{Baseline systems.} \squeeze{We compare against VStore~\cite{DBLP:conf/eurosys/XuBL19}, a recent storage system for video workloads, and direct use of the local file system. %
We build VStore with GPU support.
VStore intermittently failed when operating on ${>}2,000$ frames and so we limit 
all VStore experiments to this size.  %
}

\textbf{Experimental configuration.}  \squeeze{We perform all experiments using a single-node system equipped with an Intel i7 processor, 32GB RAM, and a Nvidia P5000 GPU.%
}

\datasetsTable

\textbf{Datasets.}
\squeezemore{We evaluate using both real and synthetic video datasets (see \autoref{table:datasets}).  We use the former to measure \name performance under real-world inputs, and the latter to test on a variety of carefully-controlled configurations.
The ``Robotcar'' dataset consists of two highly-overlapping videos from vehicle-mounted stereo cameras~\cite{RobotCarDatasetIJRR}.}  \revision{The dataset is provided as 7,494 PNG-compressed frames at 30 FPS (as is common for datasets that target machine learning).  We cropped and transcoded these frames into a \avc video with one-second GOPs.}

The ``Waymo'' dataset is an autonomous driving dataset~\cite{waymo}.  We selected one segment ($\sim$20s) captured using two vehicle-mounted cameras.  Unlike the Robotcar dataset, we estimate that Waymo videos overlap by $\sim$15\%.  %

\squeezemore{
Finally, the various ``VisualRoad'' datasets consist of synthetic video generated using a recent video analytics benchmark designed to evaluate the performance of video-oriented data management systems~\cite{DBLP:conf/sigmod/HaynesMBCC19}.  
\shepherd{To generate each dataset, we execute a one-hour simulation and produce video data at 1K, 2K, and 4K resolutions.  We modify the field of view of each panoramic camera in the simulation so that we could vary the horizontal overlap of the resulting videos.  We repeat this process to produce five distinct datasets; for example,  ``VisualRoad-1K-75\%'' has two 1K videos with 75\% horizontal overlap.
}}

\squeezemore{Because the size of the uncompressed 4K Visual Road dataset ($\sim 5$TB) exceeds our storage capacity, we do not show results that require fully persisting this dataset uncompressed on disk.}

\subsection{Data Retrieval Performance}
\label{sec:evaluation:read}

\fragmentReadFigure

\squeezemore{
\vspace{-1.25em}
\par\noindent\ignorespacesafterend 
\textbf{\revision{Long} Read Performance.} 
We first explore \name performance \revision{for large reads} at various cache sizes. %
We repeatedly execute queries of the form $read(\text{VRoad-4K-30\%}, 4\textsc{k}, [t_1, t_2], P)$, with parameters drawn at random.
We assume an infinite budget and iterate until \name has cached a given number of videos.
}

\squeezemore{
We then execute a maximal \hevc read ($t{=}[0\text{--}3600]$), which is different from the originally-written physical video (\avc).  This allows \name to leverage its materialized fragments.  %
}

\squeezemore{
\autoref{fig:read-fragments} shows performance of this read.
Since none of the other baseline systems support automatic conversion from \avc to \hevc, we do not show their runtimes for this experiment.
}

\squeezemore{As we see in \autoref{fig:read-fragments}, even a small cache improves read performance substantially---28\% at 100 entries and up to a maximum improvement of 54\%.
Further, because \name decodes fewer dependent frames, \name's solver-based fragment selection algorithm outperforms both reading the original video and a na\"ive baseline that greedily selects fragments.  
}

\textbf{\revision{Short Read Performance.}} 
\revision{
We next examine \name performance when reading small, one-second regions of video (e.g., to apply license plate detection only to regions of video that contain automobiles).  \squeeze{In this experiment, we begin with the \name state generated by the previous experiment and execute many short reads of the form $read(\text{VisualRoad-4K-30\%}, R, [t_1, t_2], P)$, where $0 \le t_1 < 3600$ and $t_2 = t_1 + 1$ (i.e., random 1~second sequences).  \shepherd{$R$ and $P$ are as in the previous experiment.}
}

\squeezemore{
\autoref{fig:small-reads} shows the result for \name (``VSS (All Optimizations)'') versus reading from the original video from the local file system (``Local FS'').  For this experiment, \name is able to offer improved performance due to its ability to serve from a cache of lower-cost fragments, rather than transcoding the source video.
We discuss the other optimizations in this plot in \autoref{sec:evaluation:compression}.
} 
}

\deferredCompressionAndSelectFragmentsFigure

\readFigure

\textbf{Read Format Flexibility.}
\squeezemore{Our next experiment evaluates \name's ability to read video data in a variety of formats.
To evaluate,
we write the VRoad-1K-30\% dataset to each system in both compressed (224MB) and uncompressed form ($\sim$328GB).
We then read video from each system in various formats and measure throughput.
\autoref{fig:read} shows read results for the same (\ref{subfig:read-same-format}) and different (\ref{subfig:read-different-format}) formats.
Because the local file system does not support automatic transcoding (e.g., \avc to \rgb), we do not show results for these cases.  Additionally, VStore does not support reading some formats; we we omit these cases.
}

\squeezemore{
We find that read performance \textit{without} a format conversion from \name is modestly slower than the local file system, due in part to 
the local file system being able to execute entirely without kernel transitions and \name's need to concatenate many individual GOPs.  However, 
\name can adapt to reads in \textit{any} format, a benefit not available when using the local file system. 
}

\squeeze{
We also find that \name outperforms VStore when reading uncompressed video and is similar when transcoding \avc.  Additionally, \name offers flexible IO format options and does not require a workload to be specified in advance.
}

\subsection{Data Persistence \& Caching}
\label{sec:evaluation:write}

\writeFigure

\textbf{Write Throughput.}
\squeezemore{
We next evaluate \name write performance by writing
each dataset to 
each
system in both compressed and uncompressed form.  For uncompressed writes, we measure throughput and show results in \autoref{subfig:write:uncompressed}.
}

\squeezemore{
For uncompressed datasets that fit on local storage, 
all systems perform similarly.
On the other hand, 
no other 
systems have the capacity to store the larger uncompressed datasets
(e.g., VisualRoad-4K-30\% is $>$5TB uncompressed).
However, 
\name's deferred compression 
allows it to store datasets no other system can handle (though with decreased throughput).
}

\squeezemore{
\autoref{subfig:write:compressed} shows results for writing the \textit{compressed} datasets to each store.
Here all perform similarly;
\name and VStore exhibit minor overhead relative to the local file system.
}

\cacheEvictionAndJointSizeFigure

\textbf{Cache Performance.}
\squeezemore{
\shepherd{
To evaluate the \name cache eviction policy, we begin by executing 5,000 random reads to populate the cache, using the same parameters as in \autoref{sec:evaluation:read}.
In this experiment, instead of using an infinite storage budget, we limit it to multiples of the input size 
and apply either the least-recently used (LRU) or \name eviction policy.}  This limits the number of physical videos available for 
reads.
With the cache populated, we execute a final read for the entire video.
\autoref{fig:eviction} plots runtimes for each policy and storage budget.
This shows that \name reduces read 
time relative to LRU.
}

\subsection{Compression Performance}
\label{sec:evaluation:compression}

\jointCompressionQualityTable

\textbf{Joint Compression Quality.}
\shepherd{
\squeezemore{
We first examine the recovered quality of jointly-compressed physical videos.  For this experiment we write various overlapping Visual Road datasets to \name.  We then  read each video back from \name and compare its quality---using peak signal-to-noise ratio (PSNR)---against its originally-written counterpart.  
}
}

\autoref{table:joint-compression-quality} gives the PSNR for recovered data compared against the written videos.
\squeeze{Recall that a PSNR of ${\ge}40$ is considered to be lossless, and ${\ge}30$ near-lossless~\cite{DBLP:conf/icpr/HoreZ10}.  
\revision{
When applying the unprojected merge function during joint compression,
\shepherd{\name achieves almost perfect recovery for the left input (with PSNR values exceeding 300dB) and near-lossless quality for the right input.}
Loss in fidelity occurs when inverting the merge, i.e., performing the inverse projection on the right frame using left-frame pixels decreases the quality of the recovered frame.  This merge function also leads to \name rejecting approximately half of the fragments due to their falling below the minimum quality threshold.  We conclude this merge function is useful for reducing storage size in video data that must maintain at least one perspective in high fidelity.}

\squeezemore{
\shepherd{
On the other hand, \name attains balanced, near-lossless quality for \textit{both} the left and right frames when applying the mean merge function during joint compression.  Additionally, the number of fragments admitted by the quality model is substantially higher under this merge function.  Accordingly, the mean merge function is appropriate for scenarios where storage size is paramount and  near-lossless degradation is acceptable.
}
}
}

\jointThroughputFigure
\jointCompressionThroughputFigure

\textbf{Joint Compression Throughput.}
We next examine read throughput with and without the joint compression optimization.  \shepherd{First, we write the VisualRoad-1K-30\% dataset to \name, once with joint compression enabled and separately with it disabled.}  We then read in various physical configurations over the full duration.   \autoref{subfig:joint-reads} shows the throughput for reads using each configuration. 
Our results indicate that read overhead when using joint compression is modest but similar to reads that are not co-compressed.

\squeeze{
Joint compression requires several nontrivial operations, and 
we next evaluate this overhead
by writing
1\textsc{k}, 2\textsc{k}, and 4\textsc{k} video and measuring throughput.  \autoref{subfig:joint-writes} shows the results.  
Joint writes are similar to writing each video stream separately.  This speedup is due to \name's encoding 
the lower-resolution streams in parallel. 
Additionally, the overhead in feature detection and generating the homography is low.
\revision{
\autoref{fig:joint-compression-throughput} decomposes joint compression overhead into these subcomponents.  \shepherd{First, \autoref{subfig:joint-compression:resolution} measures joint compression overhead by resolution, where compression costs dominate for all resolutions.}  \autoref{subfig:joint-compression:dynamic} further shows \name performance under three additional scenarios: a static camera, a slowly rotating camera that requires homography reestimation every fifteen frames, and a rapidly rotating camera that requires reestimation every five frames.  In these scenarios non-compression costs scale with the reestimation period, and compression performance is loosely correlated since a keyframe is needed after every homography change.
}
}

\squeezemore{
We next
evaluate 
\name's joint compression selection algorithm.  Using 
VisualRoad-1K-30\%, 
we count 
joint compression candidates 
using 
(i) \name's algorithm,
(ii) an 
oracle, and (iii) random selection. %
\autoref{fig:compression-selection-method} shows performance of each strategy.  \name
identifies 80\% of the 
applicable pairs in time similar to the oracle and outperforms random sampling.
}

\revision{
\shepherd{
\textbf{Joint Compression Storage.}
\squeezemore{To show the storage benefit of \name's joint compression optimization, we separately apply the optimization to each of the Visual Road videos.  We then measure the final on-disk size of the videos against their separately-encoded variants.  \autoref{fig:joint-size} shows the result of this experiment.  These results show joint compression substantially reduces the storage requirements of overlapping video.}
}
}

\textbf{Deferred Compression Performance.}
\squeezemore{
We next evaluate deferred compression for uncompressed writes by storing 3600 frames of the VisualRoad-1K-30\% dataset in \name, leaving budget and deferred compression at their defaults.}

\squeezemore{The results are listed in \autoref{fig:deferred-compression}.  \shepherd{The figure shows storage used as a percentage of the budget, throughput relative to writing without deferred compression activated, and compression level.} 
Storage used exceeds the deferred compression threshold early in the write, and a slope change shows that deferred compression is 
moderating write size.
Compression level scales linearly with storage cost.  Throughput drops substantially as compression is activated, recovers considerably, and then slowly degrades as the %
level is increased.}

\revision{
\squeezemore{
Similarly, \autoref{fig:deferred-compression-reads} shows throughput for \textit{reading} fragments of raw video compressed at various levels.  Though these reads have decreased performance and increased variance relative to uncompressed reads, at all levels ZStandard decompression remains much faster than using traditional video codecs.
}

\squeeze{Finally, \autoref{fig:small-reads} explores the trade-offs between deferred compression performance and \name's cache eviction policy.  \shepherd{In this experiment we variously disable deferred compression (``VSS (No Deferred Compression)'') and modify \name to use ordinary LRU (``VSS (Ordinary LRU)'').}  The results show that \name benefits from its eviction policy for small numbers of fragments (when deferred compression is off or at a low level) but offers increasingly large benefits as the cache grows.  At large cache sizes as the storage budget is exhausted, deferred compression is increasingly important to mitigate eviction of fragments that are subsequently useful.}
}

\phantom{XXX XXX XXX XXX XXX XXX XXX }

\subsection{End-to-End Application Performance}
\label{sec:end-to-end}

\applicationAndReadThroughputFigure

\revision{
Our final experiment evaluates the performance of the end-to-end application described in \autoref{sec:architecture}.  In this scenario, \name serves as the storage manager for an application monitoring an intersection for automobiles.  \squeeze{It involves three steps: (i) an \textit{indexing phase} that identifies video frames containing automobiles using a machine learning algorithm, (ii) a \textit{search phase} that, given an alert for a missing vehicle, uses the index built in the previous step to query video frames containing vehicles with matching colors, and (iii) a \textit{streaming content retrieval phase} that uses the search phase results to retrieve video clips containing vehicles of a given color.}

\shepherd{We implement this application using \name and a variant that reads video data using OpenCV and the local file system.  For indexing, the application identifies automobiles using YOLOv4~\cite{bochkovskiy2020yolov4} (both variants use OpenCV to perform inference using this model).  
For the search task, vehicle color is identified by
computing 
a color histogram of the region inside the bounding box.  We consider a successful detection to occur when the Euclidean distance between the largest bin and the search color is $\le 50$.  \squeeze{In the content retrieval phase, the application generates $n$ video clips by retrieving contiguous frames containing automobiles of the search color.}}

\shepherd{
\squeezemore{
We use as input four extended two-hour variants of the Visual Road 2\textsc{k} dataset.  
To simulate execution by multiple clients, we launch a separate process for each client.
Both variants index automobiles every ten frames (i.e., three times a second).  All steps exhaust all CPU resources at $>4$ clients, and so we limit concurrent requests to this maximum.
}
}

\squeezemore{
\autoref{figure:end-to-end} shows the performance of each application step.  The indexing step is a CPU-intensive operation that necessitates both video decoding and model inference, and because \name introduces low overhead for reads, both variants perform similarly.  Conversely, \name excels at executing the search step, which requires retrieving raw, uncompressed frames that were cached during the indexing step.  As such, it substantially outperforms the OpenCV variant.  Finally, \name's ability to  efficiently identify the lowest-cost transcode solution enables it to execute the streaming content retrieval step significantly faster than the OpenCV variant.  We conclude that \name's performance greatly improves end-to-end application performance for queries that depend on cached video in multiple formats, and scales better with multiple clients.
}
}

\section{Related Work}
\label{sec:related work}

\revision{
\squeezemore{
Increased interest in machine learning and computer vision has led to the development of a number of systems that target video analytics, including LightDB~\cite{DBLP:journals/pvldb/HaynesMABCC18}, VisualWorldDB~\cite{DBLP:conf/cidr/HaynesDMBCC20}, Optasia~\cite{DBLP:conf/cloud/LuCK16}, Chameleon~\cite{DBLP:conf/sigcomm/JiangABSS18}, Panorama~\cite{DBLP:journals/pvldb/ZhangK19}, Vaas~\cite{DBLP:journals/pvldb/BastaniMM20}, SurvQ~\cite{stonebraker2020surveillance}, and Scanner~\cite{DBLP:journals/tog/PomsCHF18}.  
These systems 
can be modified to leverage a storage manager like \name.
Video accelerators such as BlazeIt~\cite{DBLP:journals/pvldb/KangBZ19}, VideoStorm~\cite{DBLP:conf/nsdi/ZhangABPBF17}, Focus~\cite{DBLP:conf/osdi/HsiehABVBPGM18}, NoScope~\cite{DBLP:journals/pvldb/KangEABZ17}, Odin~\cite{DBLP:journals/pvldb/SupremAPF20}, SQV~\cite{DBLP:conf/sigmod/XarchakosK19}, MIRIS~\cite{DBLP:conf/sigmod/BastaniHBGABCKM20}, Tahoma\cite{DBLP:conf/icde/AndersonCRW19}, and Deluceva~\cite{DBLP:conf/ssdbm/WangB20} can also %
benefit from \name 
for training and inference.
}
}

\squeezemore{
Few recent storage systems target video analytics (although others have highlighted this need~\cite{DBLP:conf/hotstorage/Gupta-CledatRS17,Jiang:2019:NCN:3349614.3356026}).  
VStore~\cite{DBLP:conf/eurosys/XuBL19} 
targets machine learning workloads by staging video in pre-specified formats.  However, VStore requires a priori knowledge of the workload and 
only makes preselected materializations available.
By contrast, quFiles exploits data independence at the granularity of entire videos~\cite{DBLP:conf/fast/VeeraraghavanFNN10}.
Others have explored on-disk layout of video
for 
scalable streaming~\cite{DBLP:journals/mms/KangHW09}, and systems such as Haystack~\cite{DBLP:conf/osdi/BeaverKLSV10}, AWS Serverless Image Handler~\cite{awsimagehandler}, and VDMS~\cite{DBLP:journals/corr/abs-1810-11832} emphasize image and metadata operations.}

\revision{
Techniques similar to \name's joint compression optimization have been explored in the image and signal processing communities.  For example, Melloni et al. develop a pipeline that identifies and aligns near-duplicate videos~\cite{DBLP:conf/icip/MelloniLBTT15}, and Pinheiro et al. introduce a fingerprinting method to identify correlations among near-duplicate videos~\cite{DBLP:conf/icip/PinheiroCBTR19}.
However, unlike \name, these techniques assume that sets of near-duplicate videos are known a priori and they do not exploit redundancies to improve compression or read/write performance.
\squeeze{Finally, the multiview extension to HEVC (MV-HEVC; similar extensions exist for other codecs) attempts to exploit spatial similarity in similar videos to improve compression performance~\cite{DBLP:conf/icip/HannukselaYHL15}.  These extensions are complementary to \name, which could incorporate them as an additional compression codec for jointly-compressed video.}
} 

\squeezemore{
Finally, as in \name, the database community has long exploited data independence to improve performance.
Orthogonal optimizations could further improve \name performance (e.g., perceptual compression~\cite{DBLP:conf/cloud/MazumdarHBCCO19} or homomorphic operators~\cite{DBLP:journals/pvldb/HaynesMABCC18}).
}

\section{Conclusion}
\label{sec:conclusion}

\squeezemore{
We presented \name, a video storage system that improves the performance of video-oriented applications.  \name decouples high-level operations (e.g., machine learning) from the low-level 
plumbing to read and write data in a suitable format.  
\name automatically identifies the most efficient method to persist and retrieve video data. 
\name reduces read time by up to 54\%, and decreases the cost of persisting video by up to 45\%.
}

\revision{
\squeeze{
As future work, we plan on extending \name's joint compression optimization to support more intelligent techniques for merging overlapping pixels.  For example, \name might intelligently detect occlusions and persist both pixels in these areas.  This is important for cases where video must be maintained in its (near-)original form (e.g., for legal reasons).
}
}

\normalsize{
\squeezemore{
\par\noindent\ignorespacesafterend 
\textbf{Acknowledgments.}
This work is supported by the NSF through grants CCF-1703051, IIS-1546083, CCF-1518703, and CNS-1563788; DARPA award FA8750-16-2-0032 and RC Center grant GI18518; DOE award DE-SC0016260; a Google Faculty Research Award; an award from the University of Washington Reality Lab; a CoMotion Commercialization Fellows grant; gifts from the Intel Science and Technology Center for Big Data, Intel Corporation, Adobe, Amazon, Facebook, Huawei, and Google; and by CRISP, one of six centers in JUMP, a Semiconductor Research Corporation program sponsored by DARPA.
}
}

\bibliographystyle{ACM-Reference-Format}
\bibliography{header,self,others}

\end{sloppypar}

\end{document}